\documentclass[useAMS]{mn2e}

\usepackage{graphicx,color}
\usepackage{latexsym}
\usepackage{multirow}
\usepackage{amsmath,amssymb}        
\usepackage[draft=false]{hyperref}

\title[Radial accretion of dark matter]{Exploring the effects of pressure on the radial accretion of dark matter by a Schwarzschild supermassive black hole}

\author[F. S. Guzm\'an, F. D. Lora-Clavijo]{F. S. Guzm\'an, F. D. Lora-Clavijo\thanks{E-mail:
guzman@ifm.umich.mx (FSG); fadulora@ifm.umich.mx (FDLC)} \\
	     Instituto de F\'{\i}sica y Matem\'{a}ticas, Universidad
              Michoacana de San Nicol\'as de Hidalgo.\\ Edificio C3, Cd.
              Universitaria, 58040 Morelia, Michoac\'{a}n,
              M\'{e}xico.}

\begin{document}

\date{\today}

\pagerange{\pageref{firstpage}--\pageref{lastpage}} \pubyear{2011}

\maketitle

\label{firstpage}


\begin{abstract}
Based on the numerical solution of the time-dependent relativistic Euler equations onto a fixed Schwarzschild background space-time, we estimate the accretion rate of radial flow toward the horizon of a test perfect fluid obeying an ideal gas equation of state.  We explore the accretion rate in terms of the initial density of the fluid for various values of the inflow velocity in order to investigate whether or not sufficiently arbitrary initial conditions allow a steady state accretion process depending on the values of the pressure. We extrapolate our results to the case where the fluid corresponds to dark matter and the black hole is a supermassive black hole seed. Then we estimate the equation of state parameters that provide a steady state accretion process. We found that when the pressure of the dark matter is zero, the black hole's mass grows up to values that are orders of magnitude above $10^{9}M_{\odot}$ during a lapse of 10Gyr, whereas in the case of the accretion of the ideal gas dark matter with non zero pressure the accreted mass can be of the order of $\sim 1M_{\odot}/10Gyr$ for black holes of $10^{6}M_{\odot}$. This would imply that if dark matter near a supermassive black hole acquires an equation of state with non trivial pressure, the contribution of accreted dark matter to the supermassive black hole growth could be small, even though only radial accretion is considered.
\end{abstract}




\section{Introduction}

The problem of formation and evolution of supermassive black holes (SMBHs) in the center of gigant elliptic and disk galaxies remains unsolved. The black hole growth is usually related to its coexistence with the surrounding matter, both baryonic and dark matter. Some models consider these black holes are the result of the evolution of seed black holes of various initial masses that grow through accretion, for instance, if there are seed black holes as the result of primordial gas clouds collapse at the early evolution of galaxies, the seeds would have masses of the order of $10^3$ to $10^4 M_{\odot}$ \cite{Eisenstein1995,Bullock2004}, whereas when the seeds are considered to be the result of the collapse of relativistic massive stars the seeds should have started with masses of the order of hundreds of solar masses \cite{Heger2003}; also combined models of the collapse of primordial gas and dark matter are assumed as black hole seeds of intermediate masses \cite{Umeda2009}. The study of bolometric quasar luminosity function at various distances indicates that the accreted mass of supermassive black holes should be baryonic \cite{Small1992,Hopkins2007}, however no final answer about the amount of accreted dark matter by SMBHs is available, for instance in \cite{Peirani2008} it is shown that dark matter contributes with at most $10\%$ of the total accreted mass, which together with observations of the bolometric quasar luminosity confirms that baryons are the main matter component that feeds the black hole through accretion.

Standard analyses of SMBHs growth are based on the study of the evolution of phase space distributions, for instance, if SMBHs are primarily fed by collisionless dark matter, or stars, the key problem to overcome is how to refill the loss cone \cite{LighShap1977,ZhaoRees2002}. Previous work has shown that the timescale for this is too long for collisionless matter to contribute significantly to black hole growth \cite{ReadGilmore2003}, though box orbits in a triaxial potential might provide an interesting caveat to this argument \cite{Holley2006}. Another possibility is runway growth at the center of dense star clusters. This could produce intermediate mass black holes, which would be seed black holes that grow through later gas accretion \cite{Ebisuzaki2001,Zwart2002}, which is expected to be fueled by mergers \cite{Volonteri2005}. Concerning the non-radial accretion process, particles have to overcome the angular momentum barrier in order to get the gas to the very center of a galaxy \cite{King2006}. Mechanisms proposed include the "bars within bars" model \cite{Shlosman1989,Begelman2006} and direct collapse of cold clouds \cite{Loeb1994}.

On the other hand, non standard models consider the SMBHs grow through the direct accretion of self-interacting or collisional dark matter \cite{Ostriker2000, Hu2006}. In this case, the self interaction cross section is tuned to ensure that the loss cone is refilled by diffusion on an interestingly short timescale. Collisional dark matter has also been studied at galactic scale as a possible solution to the galactic core problem \cite{Spergel2000} and the corresponding gravitational collapse \cite{Moore2000}. The collisional dark matter also finds its candidates from extensions to the standard model of particles, for example singlets \cite{HolzZee2001}.

In this paper we present a totally different approach that considers both the collisional and the collisionless dark matter. It consists in the solution of the fully relativistic time-dependent Euler equations for a fluid on a Schwarzschild black hole space-time, where the fluid plays the role of dark matter. We model the collisional and collisionless  dark matter as a relativistic ideal gas with and without pressure respectively, in order to study the possibility that the pressure itself suffices to stop the accretion and pull the system toward a stationary state that allows the SMBH to have the observed masses. We measure the accretion rate of dark matter into the black hole parametrized by the equation of state, initial dark matter density and initial infall velocity.

Given that we only consider radial accretion, our results for both the collisional and collisionless cases are upper bounds to more general accretion situations where non-trivial impact parameters of the dark fluid particles are considered, for instance those where rotation of dark matter particles is considered.

The paper is organized as follows. In section \ref{sec:euler} we define the relativistic Euler equations describing the fluid on a Schwarzschild black hole, whereas in section \ref{sec:numerics} we explain the numerical methods we use; in section \ref{sec:results} we show our results for the extrapolation to the supermassive black hole plus dark matter system and in  \ref{sec:conclusions} we summarize our  conclusions.


\section{Relativistic Euler's equations}
\label{sec:euler}

\subsection{In general}

In order to track the evolution of a perfect fluid on a non-flat space-time it is necessary to write down the general relativistic Euler equations. For a generic fixed space-time these can be derived from the local conservation of the stress-energy tensor $T^{\mu \nu}$ and the local conservation of the number of baryons:

\begin{eqnarray}
 \nabla_{\nu}(T^{\mu \nu})=0, \label{eq:conserv1}\\
 \nabla_{\mu}(\rho u^{\mu}) = 0, \label{eq:conserv2}
\end{eqnarray}

\noindent where $\rho$ is the proper rest mass density, $u^{\mu}$ is the 4-velocity of the fluid and $\nabla_{\mu}$ is the covariant derivative consistent with the 4-metric of the space-time \cite{mtw}.

The stress energy tensor representing a perfect fluid in General Relativity is given by

\begin{equation}
T^{\mu \nu} = \rho h u^{\mu}u^{\nu}+p g^{\mu \nu}. \label{eq:emt}
\end{equation}

\noindent where $p$, $h$ and $g^{\mu \nu}$ are respectively the pressure, the specific enthalpy and the 4-metric tensor of the space-time, respectively. 

We choose to write down the space-time metric in the 3+1 decomposition fashion for a coordinate system $(t,x^i)$, that is \cite{mtw}:

\begin{equation}
ds^2 = -(\alpha^2 - \beta_i \beta^i)dt^2 + 2\beta_i dx^i dt + \gamma_{ij}dx^i dx^j, \label{eq:lineelement}
\end{equation}

\noindent where the latin indices $i,j$ label the three spatial coordinates of the space-time, $\alpha$ is called the lapse function and together with the shift vector $\beta^i$ determine the foliation of the space-time in space-like hypersurfaces; $\gamma_{ij}$ is the induced three metric that relates proper distances on the spatial hypersurfaces \cite{mtw}. 

With this information, the unknown quantities of Euler equations are the following primitive variables: the rest mass density of the fluid $\rho$, its pressure $p$ and its velocity measured by an eulerian observer $v^i$ defined in terms of the spatial part of the 4-velocity of the fluid by $v^i = \frac{u^i}{W} + \frac{\beta^i}{\alpha}$, where $W$ is the Lorentz factor $W=\frac{1}{\sqrt{1-\gamma_{ij} v^i v^j}}$. Given that there are five unknowns and four equations the system is closed with an equation of state $p=p(\rho,\epsilon)$, where $\epsilon$ is the specific internal energy, defined by $h=1+\epsilon+p/\rho$.

Even though one can write down relativistic Euler equations straightforwardly in terms of the primitive variables, we  work with conservative variables that allow one to write down the relativistic Euler equations as a set of flux balance laws

\begin{equation}
\frac{\partial {\bf u}}{\partial t} + \frac{\partial {\bf F}({\bf u})^{i}}{\partial x^i}= {\bf S},
\end{equation} 

\noindent where ${\bf q}$ is the vector of conservative variables, ${\bf F^{i}}$ are the fluxes in the three spatial directions, ${\bf S}$ is a source vector. This approach is adequate to implement the finite volume methods described below. For the perfect fluid above, a flux balance system of equations is \cite{valencia1991,valencia1994,valencia1997,wai-mo}:

\begin{eqnarray}
&& {\bf u} = \left [
\begin{array}{c}
D \\
J_j \\
\tau 
\end{array}
\right] =
\left [
\begin{array}{c}
\sqrt{\gamma} \rho W \\
\sqrt{\gamma} \rho h W^2 v_j \\
\sqrt{\gamma}(\rho h W^2 -p - \rho W)
\end{array}
\right] , \label{eq:qvector} \\ \nonumber
\\ 
&&{\bf F}^ i= 
\left [
\begin{array}{c}
\alpha \left (v^i - \frac{\beta^i}{\alpha} \right )D \\
\alpha \left (v^i - \frac{\beta^i}{\alpha} \right )J_j + \alpha \sqrt{\gamma} p \delta_j^i\\
\alpha \left (v^i - \frac{\beta^i}{\alpha} \right )\tau + \alpha \sqrt{\gamma} v^i p 
\end{array}
\right] , \label{eq:fvector} \\ \nonumber
\\
&&{\bf S} = 
\left [
\begin{array}{c}
0\\
\alpha \sqrt{\gamma} T^{\mu \nu} g_{\nu \sigma} \Gamma^{\sigma}_{\mu j} \\
\alpha \sqrt{\gamma} (T^{\mu 0} \partial_{\mu}\alpha - \alpha T^{\mu \nu} \Gamma^{0}_{\mu \nu})
\end{array}
\right] , \label{eq:svector}
\end{eqnarray}

\noindent where $\gamma=\det(\gamma_{ij})$ is the determinant of the 3-metric and $\Gamma^{\sigma}{}_{\mu\nu}$ are the Christoffel symbols.

\subsection{Equations for the spherically symmetric case}

We now specialize to the case of a Schwarzschild black hole, for which the line element in Eddington-Finkelstein coordinates $x^{\mu}=[t,r,\theta,\phi]$,  is written for the line element (\ref{eq:lineelement}) with the following gauge and 3-metric

\begin{eqnarray}
&& \alpha = \frac{1}{\sqrt{1 + \frac{2M}{r}}}, \nonumber\\
&& \beta^i=\left [ \frac{2M}{r}\left(\frac{1}{1+\frac{2M}{r}}\right),0,0\right],  \nonumber\\
&& \gamma_{ij}=diag[1+\frac{2M}{r},r^2,r^2\sin^2\theta],
\end{eqnarray}

\noindent where $M$ is the mass of the black hole. 

Notice that the radial is the only non-trivial component of the shift vector $\beta^r$ and also the only non-zero velocity is $v^r$. This implies that there are only three non-trivial balance law equations.
In terms of these geometric quantities, and provided the space-time is spherically symmetric, the system of flux balance equations reads:

\begin{eqnarray}
{\bf u} &=& 
	\left[
	\begin{array}{c}
	D\\ 
	J_r \\ 
	\tau 
	\end{array}
	\right], \nonumber\\
{\bf F}^r &=& 
	\left[
	\begin{array}{c}
	\alpha \left(v^r - \frac{\beta^r}{\alpha} \right)D \\
	\alpha\left( v^r - \frac{\beta^r}{\alpha} \right)J_r + \alpha \sqrt{\gamma} p\\
	\alpha \left( v^r - \frac{\beta^r}{\alpha}\right)\tau + \sqrt{\gamma}\alpha v^r p
	\end{array}
	\right],\nonumber\\
{\bf S} &=&
	\left[
	\begin{array}{c}
	0\\
	\alpha \sqrt{\gamma}T^{\mu\nu}g_{\nu\sigma}\Gamma^{\sigma}{}_{\mu r}\\
	\alpha \sqrt{\gamma} (T^{\mu 0}\partial_{\mu} \alpha - \alpha T^{\mu\nu}\Gamma^{0}{}_{\mu\nu})
	\end{array}
	\right]\label{eq:sphericalConservative}
\end{eqnarray}

\noindent which obey the balance law

\begin{equation}
\frac{\partial {\bf u}}{\partial t} + \frac{\partial {\bf F}^r({\bf u})}{\partial r} = {\bf S}.\label{eq:FluxConservativeSpherical}
\end{equation}

Finally, we adopt an ideal gas equation of state, in which the pressure is given in terms of the rest mass density and specific internal energy by

\begin{equation}
p = (\Gamma - 1)\rho\epsilon. \label{eq:eos}
\end{equation}

\noindent where $\Gamma$ is the adiabatic index or the ratio of specific heats.


\section{Numerical methods}
\label{sec:numerics}

{\it Solution of the equations.} In order to solve the system of equations (\ref{eq:sphericalConservative}) and (\ref{eq:FluxConservativeSpherical}) we provide initial data for the rest mass density and velocity of the fluid, and use the equation of state (\ref{eq:eos}) to calculate the initial pressure in case it is non zero. In all our production runs we use a constant density profile as initial data and various values of the constant density and velocity.

We evolve the initial data using a finite volume approximation of equations (\ref{eq:sphericalConservative}) together with High Resolution Shock Capturing methods involving approximate Riemann solvers that are able to track the potential shocks formed during the evolution of  relativistic fluid equations \cite{LeVeque}. We specially use the HLL Riemann solver and a constant piece-wise reconstruction of variables at the inter cell boundaries \cite{Toro}.

We also implement an atmosphere that prevents the rest mass density to be small enough as to provoke that relativistic Euler equations diverge or develop unphysical results, because when the density is extremely small the enthalpy diverges which provokes the numerical implementation to fail, which combines  with round off errors through arithmetic operations, cumulate during the evolution and finally produce approximations that are not accurate. The atmosphere we use is such that $\rho = \max(\rho,10^{-10})$, which is a value that allows the convergence of our numerical methods. The need of the atmosphere is in fact one of the reasons why it is not -at the moment- trivial to simulate evolutions of fluids with ultra low densities and the reason why our results on dark matter are actually extrapolations of numerical experiments with rather higher densities of matter.

{\it Boundaries and boundary conditions.} We carry out our simulations on a finite domain $0 < r_{exc} \le r \le r_{max}$. On the one hand, given that we use Eddington-Finkelstein coordinates, which penetrate the horizon, we apply the excision technique \cite{SeidelSuen1992}, that is, a chunk of the domain well inside the black hole's horizon is removed from the numerical domain from $0< r \le r_{exc}$ in order to avoid the black hole's singularity and the steep gradients of metric functions near there; since the light cones inside the horizon point toward the singularity there is no need to impose boundary conditions at $r=r_{exc}$, instead, the fluid simply gets out the domain  toward the singularity through such boundary. We choose the excision radius to be at $r_{exc}=M$, that is, half the Schwarzschild radius, which is both, far enough from the singularity at $r=0$ and provides a buffer zone $M < r < 2M$ for the fluid inside the horizon to flow smoothly from the horizon toward the excised chunk of the domain.

On the other hand, we implement an artificial outer boundary at a finite distance from the black hole at $r=r_{max}$, and apply simple boundary conditions through the extrapolation of the conservative variables. We have found that such boundary condition makes the artificial boundary to behave appropriately, that is, we have verified that our calculations do not depend on the location of the external boundary, which is the requirement that must be fulfilled by a boundary boundary condition.

{\it Diagnostics.} In order to diagnose our simulations on the fly we implement detectors located at various radii in the numerical domain, that is, we define spheres where we calculate scalar quantities. In particular we  track the accretion mass rate as a function of time. In order to do so we calculate the accretion mass rate specialized for spherical flow on a spherically symmetric space-time by 

\begin{equation}
\dot{M}_{acc} = -4 \pi r^2  \rho W \left( v^r - \frac{\beta}{\alpha} \right),
\end{equation}

\noindent at various spherical surfaces, including the black hole's event horizon.

Since we are very interested in checking the validity of our numerical results we also monitor both, the conservative and the primitive variables of our system of equations in the whole numerical domain. Then   we apply a self-convergence test of our variables in order to verify that our calculations carried out under a discretized domain converge to a correct solution in the continuum limit. That is, we want to make sure that our implementation is consistent, which in practice means that when increasing the numerical resolution, our numerical solution converges fast enough toward a continuous solution of the continuous  equations, which is a highly non-trivial test \cite{LeVeque}.

{\it Units.} 
We use geometrical units $G=c=1$, and time and spatial units of $M$; a typical dark matter density $\rho_{DM} \sim 100 M_{\odot}/pc^3$ corresponds to $\rho_{DM} \sim 10^{-26}M^{-2}$ to $10^{-30}M^{-2}$ for black hole seed masses of $10^6$ and $10^4M_{\odot}$ respectively. However, due to numerical restrictions in the use of ultralow fluid densities -as corresponds to dark matter- in geometrical units when solving the relativistic Euler equations; we study the accretion of fluid densities numerically tractable and then extrapolate our results to the density corresponding to dark matter and extract our conclusions.


\section{Results}
\label{sec:results}

\subsection{Preparing the initial conditions}

In order to solve Euler equations (\ref{eq:sphericalConservative}) and (\ref{eq:FluxConservativeSpherical}) we have to provide initial data at $t=0$, for $\rho$ and $v^r$ that for each value of $\Gamma$ would imply different values for $p$. In principle there is no prescription for $\rho$ and $v^r$ which would be arbitrary. Since we are interested in searching steady state scenarios we are inclined to set rather unbiased out of equilibrium initial conditions in order to see whether or not the system relaxes and approaches a sort of late-time attractor solution.

We have found that an initial constant density profile corresponds to a non-steady state condition, and instead produces an immediate dynamical behavior. We also choose different values of $v^r$, because we want to explore the space parameter in this direction too. We have found that the initial radial velocity is only restricted in the sense that for big initial inward velocities the fluid may reach ultra relativistic speeds when approaching the event horizon. Finally, the pressure is obtained from the equation of state.

An initial parameter for which there is no prescription is that of the internal energy of the gas $\epsilon$ in the equation of state. Its initial value determines the pressure at the initial time with the property that when $\epsilon$ approaches zero the initial conditions corresponds to the $p=0$ case.

\subsection{Case $p=0$}

\begin{figure}
\includegraphics[width=4cm]{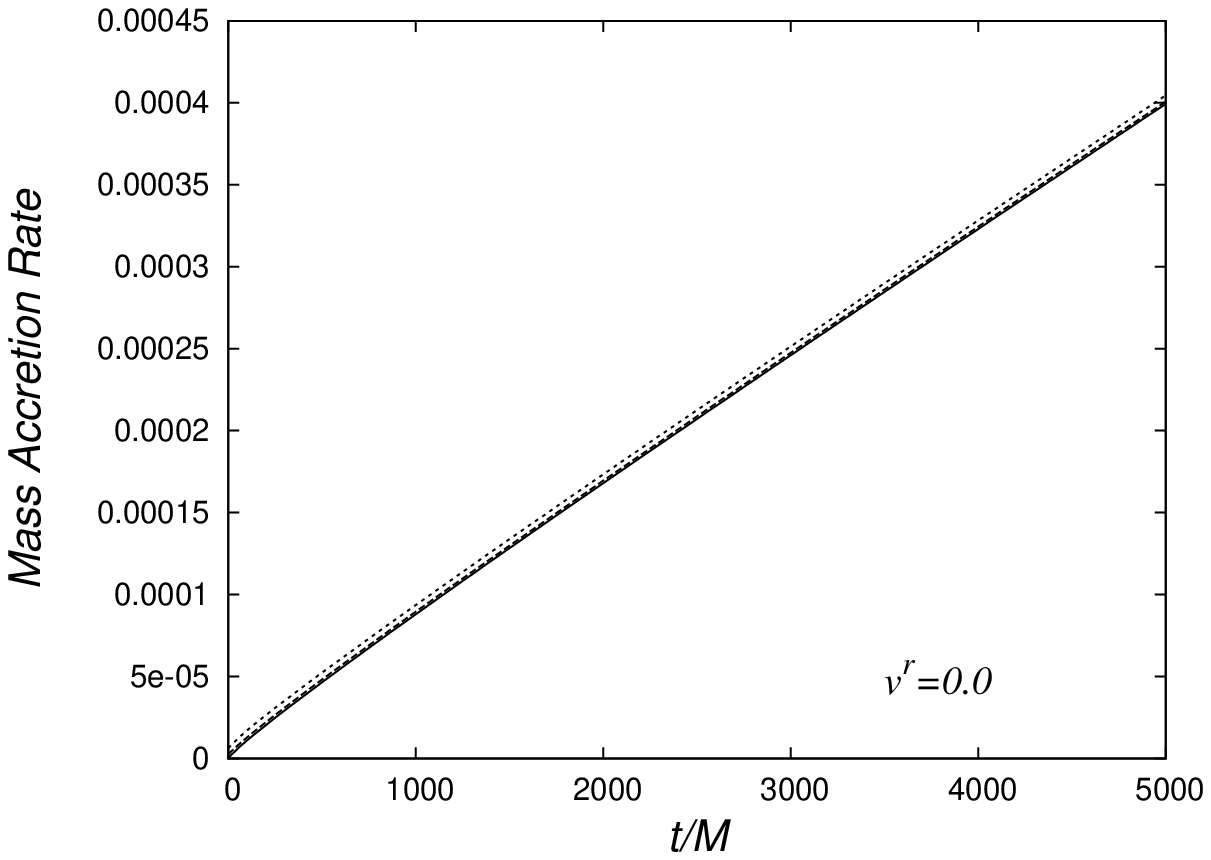}
\includegraphics[width=4cm]{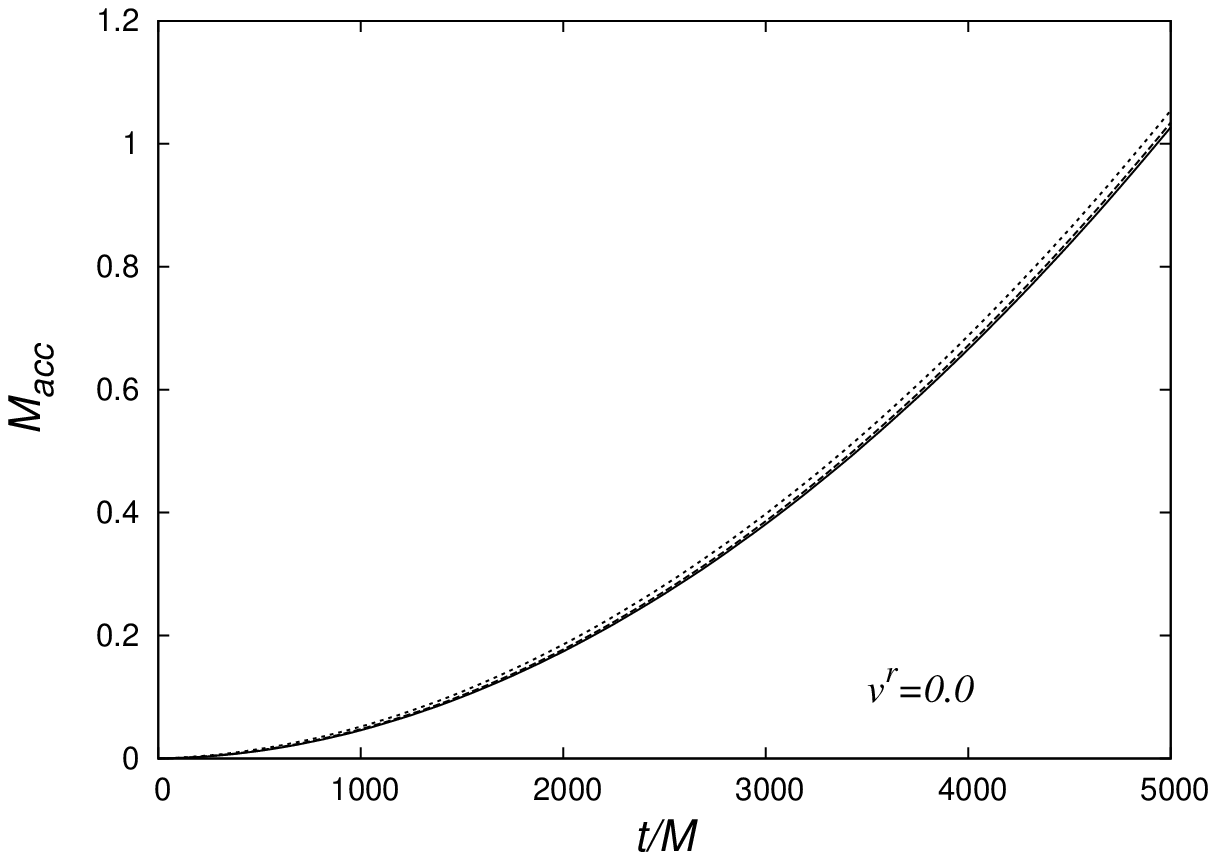}
\includegraphics[width=4cm]{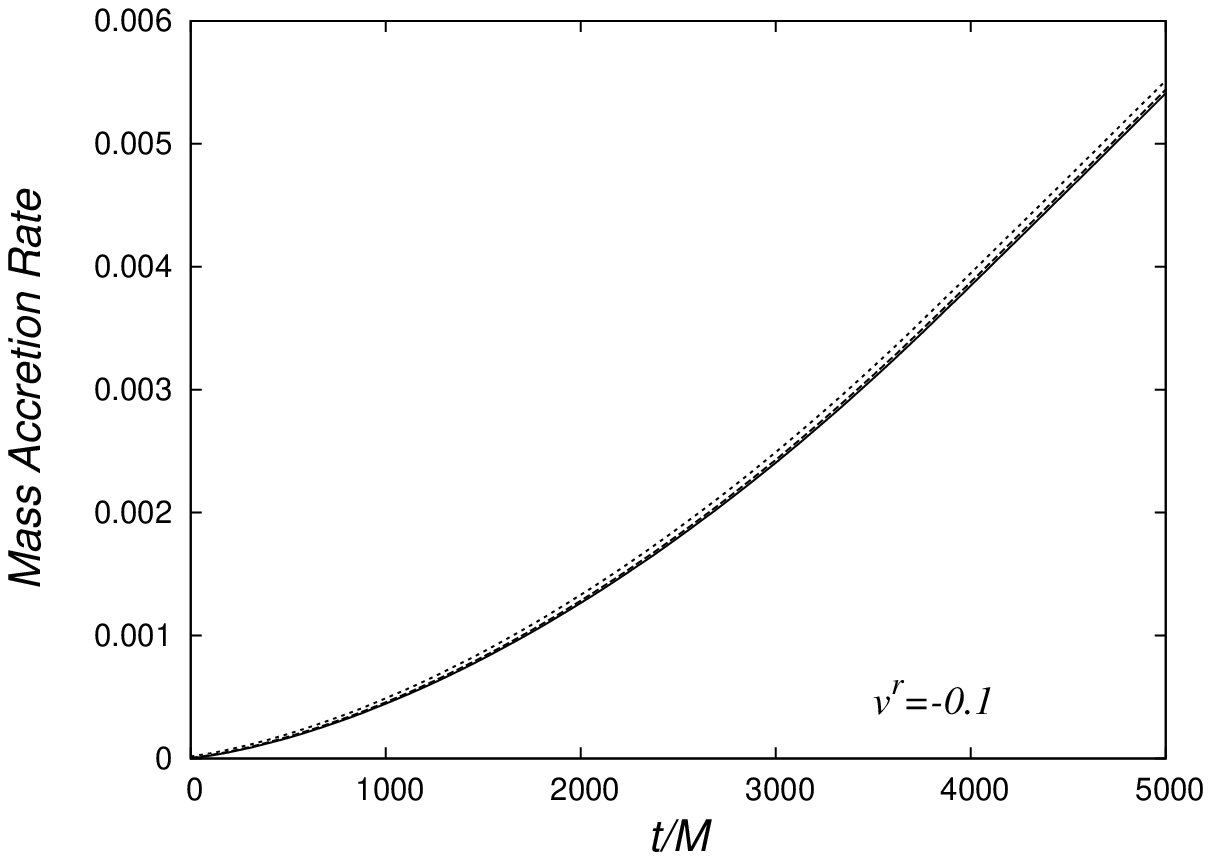}
\includegraphics[width=4cm]{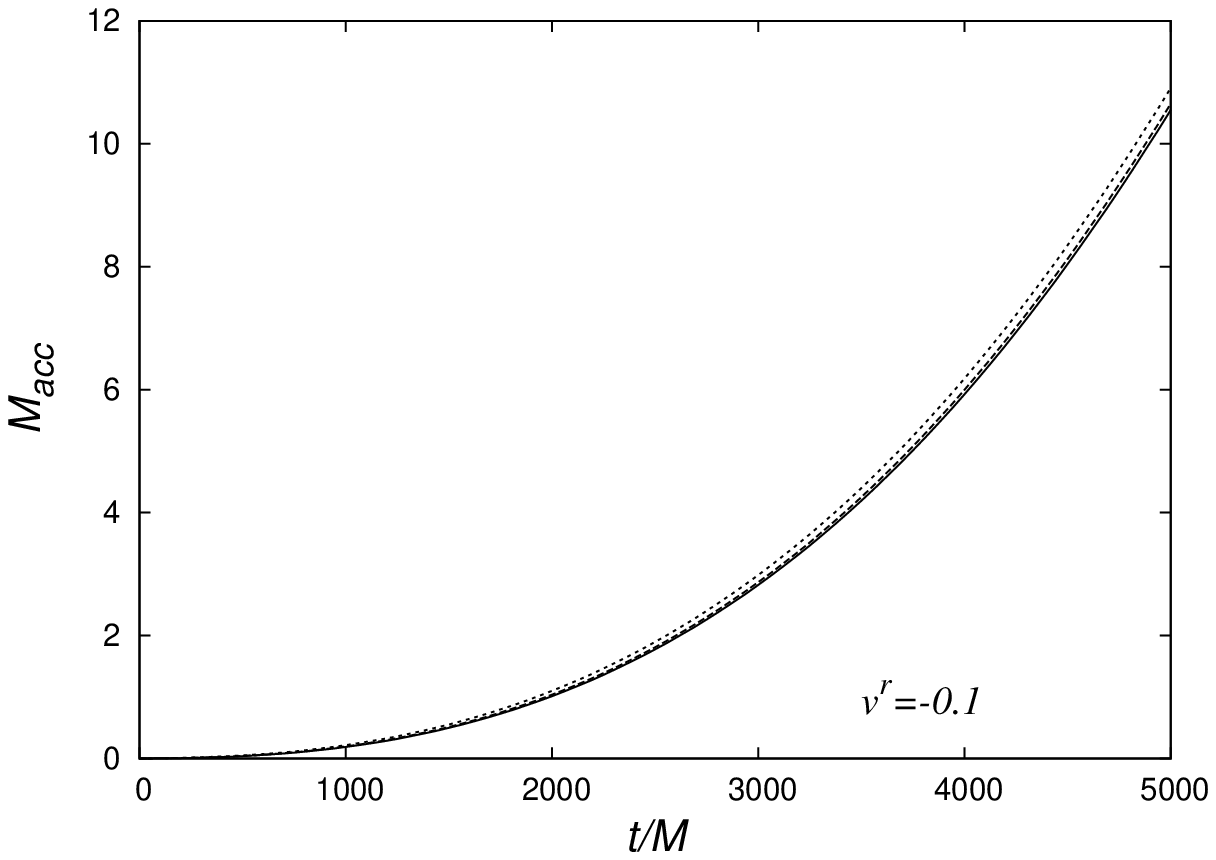}
\includegraphics[width=4cm]{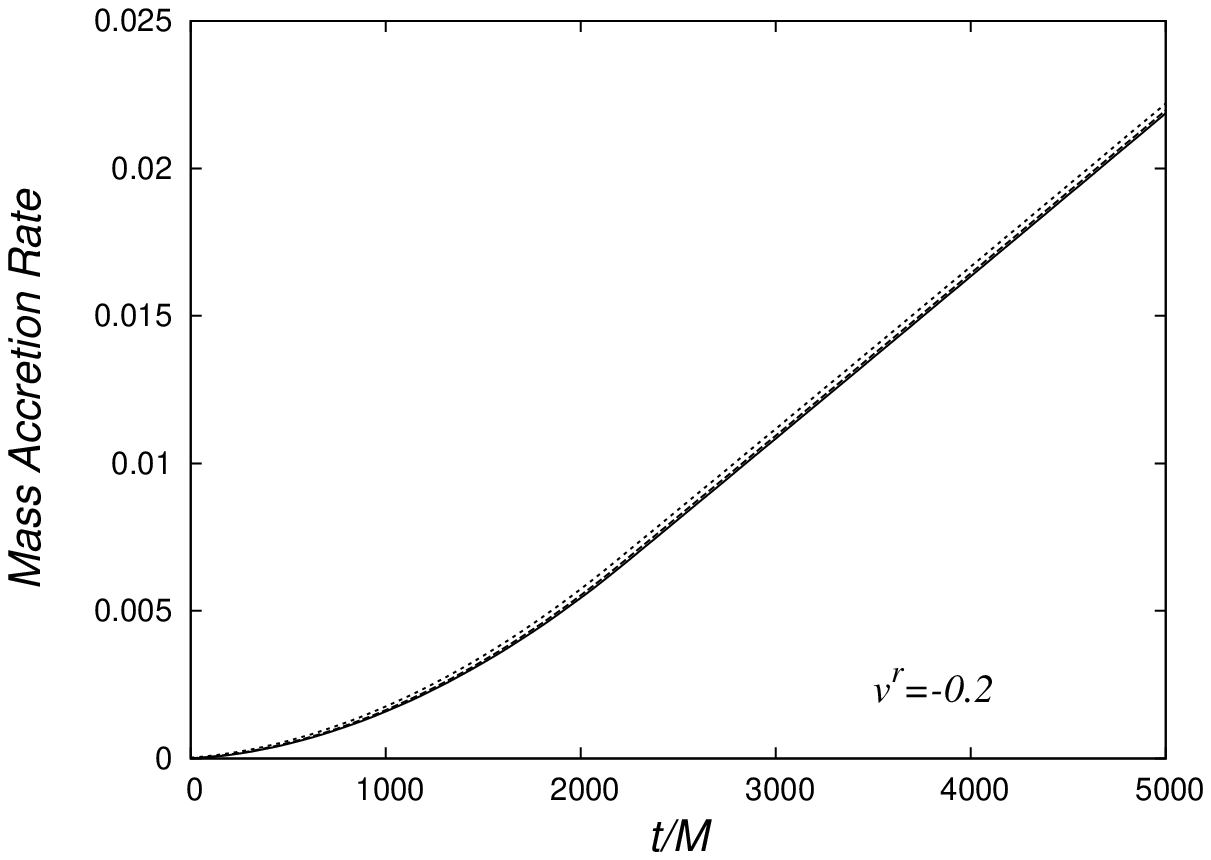}
\includegraphics[width=4cm]{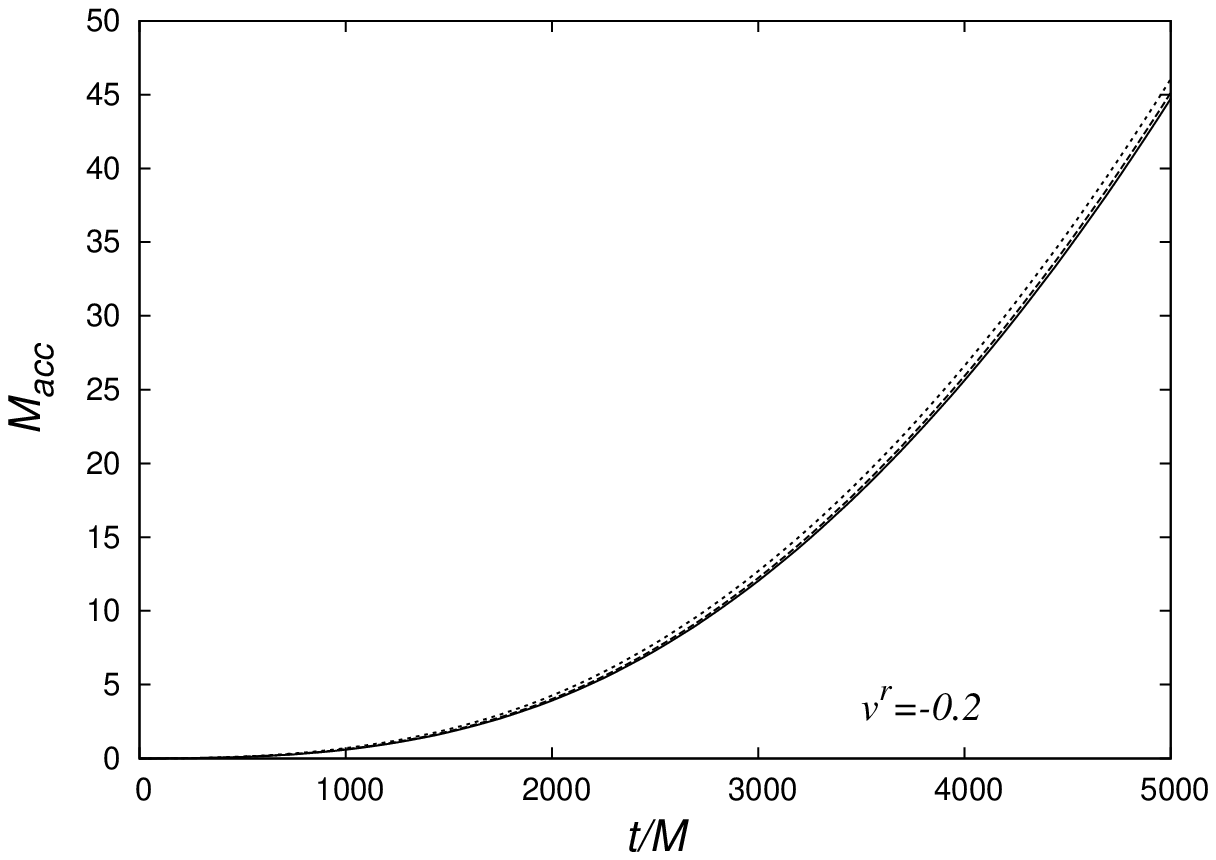}
\includegraphics[width=4cm]{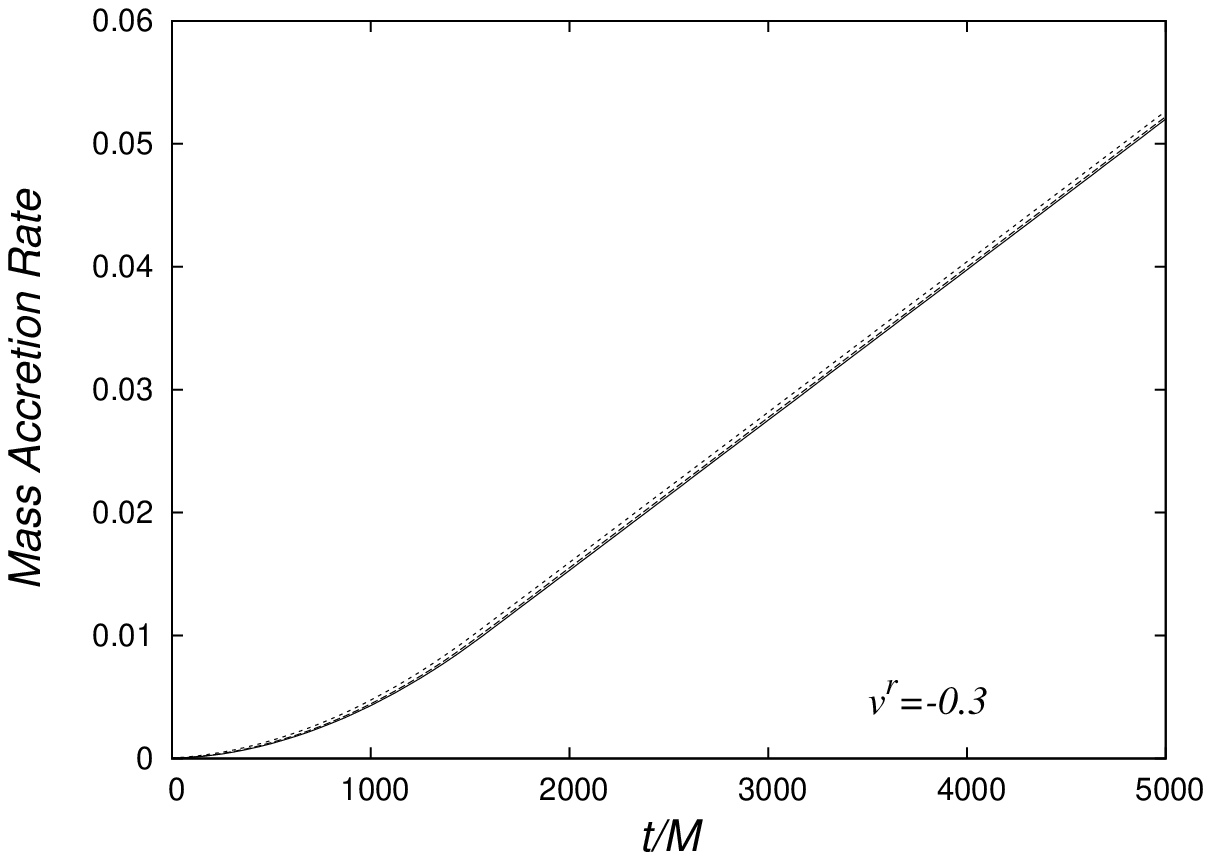}
\includegraphics[width=4cm]{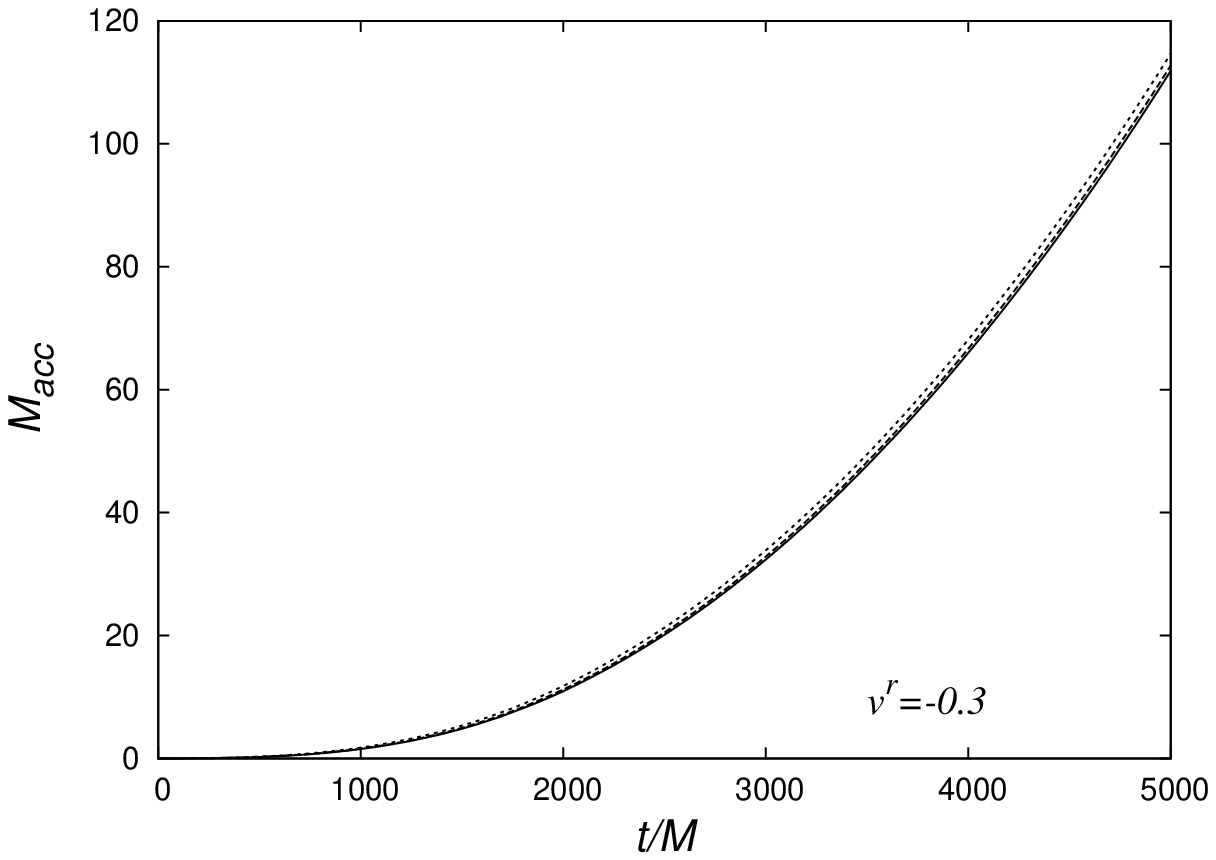}
\caption{\label{fig:MassAccretRate}  In this figure we present the accretion mass rate and the accreted mass $M_{acc}$ for a pressure-less fluid  for various values of the initial coordinate velocity $v^r=0.0,-0.1,-0.2,-0.3$. In all the cases, the initial density used is $\rho=10^{-8}$. The different types of line indicate that the accretion is being measured at various spherical surfaces, and the fact that they behave similarly shows that the trend of the accretion rate is consistent at the various detectors.}
\end{figure}

In Fig. \ref{fig:MassAccretRate}, we show the numerical results of the mass accretion rate and the accreted mass for different values of the initial inward velocity  with an initial constant density profile $\rho=10^{-8}$;  the accretion rate is measured at three different surfaces located at $r=2M,14M,29M$ in order to make sure the trend of the accretion rate is not an artifact due to backscattering near the horizon.

In all cases the mass accretion rate increases rapidly and never during our evolutions enters in a steady state. In order to measure how fast the accreted mass grows we fit the accreted mass with an anzats $M = At^B$ from our numerical results. We also extrapolate the parameters $A$ and $B$ to the cases of two reasonable density values corresponding to dark matter, and black hole seeds of $10^4$ and $10^6$ solar masses. We summarize these results in Table \ref{tab:MARDust}, where we show the values of $A$ and $B$ that better fit our numerical results, where the last two columns are an extrapolation corresponding to densities of $100 M_{\odot}/pc^3$ around black holes of $10^4$ and $10^6M_{\odot}$. The domain we use in these numerical experiments is $M<r<501M$ and made sure the running time $t=5000M$ is adequate for this grid size and our code converges.

In order to explore the possibility that a pressure less fluid can be the dark matter we consider two particular experiments:

\begin{itemize}
\item[i)] we choose the system to start accreting 10Gys ago, with an initial dark matter density of $100 M_{\odot}/pc^3$ onto a black hole with mass $10^4 M_{\odot}$,
\item[ii)] and the second system starts accreting also 10Gys ago onto a black hole seed of $10^6 M_{\odot}$,
\end{itemize} 

\noindent for which the fitting parameters are available in the last two columns of Table \ref{tab:MARDust}. The results for these two experiments are presented in Table \ref{tab:MassToday}, where it is shown that the black holes should have accreted enough dark matter as to achieve a current mass of $10^{11}M_{\odot}$ in the most conservative cases or even $10^{22}M_{\odot}$ in the worst case. We remind the reader that given that we only consider radial accretion, these results work fine as upper bounds for more general symmetries involving non-trivial impact parameters of the volume elements of the dark fluid.

It can be noticed that the initial velocities explored range from zero up to about one third the speed of light, however we found similar results for even higher speeds nearly half the speed of light. 

\begin{table*}
\begin{minipage}{180mm}
\caption{\label{tab:MARDust}In this table we show the fit parameters describing the mass growth function for a wide set of initial density and inward velocity of the test fluid for the pressure-less case. We also show the extrapolation value corresponding to the dark matter density $\rho = 10^{-26}$ and $\rho=10^{-30}$ in geometric units, that corresponds two particular cases we analyze below.}
\resizebox{7.0in}{!}{
\begin{tabular}{|l|l|l|llllllllllll} \hline
\multicolumn{9}{|c|}{Extrapolating the accreted mass (Dust Fluid)} \\ \hline
\multirow{1}{*}{} 
& $\rho=10^{-2}$ & $\rho=10^{-4}$ & $\rho=10^{-6}$ & $\rho=10^{-8}$ & $\rho=10^{-10}$ (Floor) & . . .& $\rho=10^{-26}$ & . . .& $\rho=10^{-30}$\\ \hline
\multirow{2}{*}{$v^r = 0.0$} 
& $A=0.062506 \pm 0.000103  $ & $ A=(0.062506 \pm 0.000103)*10^{-2} $ & $ A=(0.062506 \pm 0.000103)*10^{-4} $ & $ A=(0.062506 \pm 0.000103)*10^{-6} $ & $ A=(0.062506 \pm 0.000103)*10^{-8} $ & . . .& $ A=(0.062506 \pm 0.000103)*10^{-24} $ & . . .& $ A=(0.062506 \pm 0.000103)*10^{-28} $ \\
& $B=1.9506    \pm 0.0001769 $ & $ B=1.9506    \pm 0.0001769            $ & $ B=1.9506    \pm 0.0001769            $ & $ B=1.9506    \pm 0.0001769             $ & $ B=1.9506    \pm 0.0001769            $ & . . .& $ B=1.9506    \pm 0.0001769              $  & . . .& $ B=1.9506    \pm 0.0001769 $\\
\hline
\multirow{2}{*}{$v^r = -0.1$}
& $ A=0.00423575 \pm 8.719*10^{-5} $ & $ A=(0.00423575 \pm 8.719*10^{-5})*10^{-2} $ & $ A=(0.00423575 \pm 8.719*10^{-5})*10^{-4} $ & $ A=(0.00423575 \pm 8.719*10^{-5})*10^{-6} $ & $ A=(0.00423575 \pm 8.719*10^{-5})*10^{-8} $ & . . . & $ A=(0.00423575 \pm 8.719*10^{-5})*10^{-24} $ & . . . & $ A=(0.00423575 \pm 8.719*10^{-5})*10^{-28} $ \\
& $ B=2.53978      \pm 0.002341       $ & $ B=2.53978      \pm 0.002341                   $ & $ B=2.53978       \pm 0.002341                   $ & $ B=2.53978      \pm 0.002341                   $ & $ B=2.53978       \pm 0.002341                  $ & . . . & $ B=2.53978       \pm 0.002341                    $ & . . . & $ B=2.53978       \pm 0.002341                    $ \\
\hline
\multirow{2}{*}{$v^r = -0.2$} 
& $A=0.173486 \pm 0.006026   $ & $ A=(0.173486 \pm 0.006026)*10^{-2} $ & $ A=(0.173486 \pm 0.006026)*10^{-4} $ & $ A=(0.173486 \pm 0.006026)*10^{-6} $ & $ A=(0.173486 \pm 0.006026)*10^{-8} $ & . . .& $ A=(0.173486 \pm 0.006026)*10^{-24} $ & . . .& $ A=(0.173486 \pm 0.006026)*10^{-28} $ \\
& $B=2.27724   \pm 0.003729   $ & $ B=2.27724   \pm 0.003729             $ & $ B=2.27724   \pm 0.003729             $ & $ B=2.27724   \pm 0.003729          $ & $ B=2.27724   \pm 0.003729          $ & . . .& $ B=2.27724   \pm 0.003729          $ & . . .& $ B=2.27724   \pm 0.003729          $ \\
\hline
\multirow{2}{*}{$v^r = -0.3$} 
& $A=0.832518 \pm 0.02417   $ & $ A=(0.832518 \pm 0.02417)*10^{-2} $ & $ A=(0.832518 \pm 0.02417)*10^{-4} $ & $ A=(0.832518 \pm 0.02417)*10^{-6} $ & $ A=(0.832518 \pm 0.02417)*10^{-8} $ & . . . & $ A=(0.832518 \pm 0.02417)*10^{-24} $ & . . . & $ A=(0.832518 \pm 0.02417)*10^{-28} $ \\
& $B=2.19968   \pm 0.00342   $ & $ B=2.19968   \pm 0.00342          $ & $ B=2.19968   \pm 0.00342           $ & $ B=2.19968   \pm 0.00342         $ & $ B=2.19968   \pm 0.00342         $ & . . . & $ B=2.19968   \pm 0.00342           $  & . . . & $ B=2.19968   \pm 0.00342           $ \\
\hline
\end{tabular} }
\end{minipage}
\end{table*}

\begin{table}
\caption{\label{tab:MassToday} In this table, we show the mass accreted by the a black hole since 10Gys ago to the present for two black hole seeds. The initial density profile of the dark matter is $\rho=100 M_{\odot} pc^{-3}$, which in geometric units, for black hole seed masses  $M_{(i)}=10^{4}M_{\odot}$ and $M_{(ii)}=10^{6}M_{\odot}$,  correspond to $\rho = 1.17\times10^{-30}$ and $\rho = 1.17\times10^{-26}$ in geometric units respectively. The values of the mass accreted for these densities are taken from the extrapolation in Table \ref{tab:MARDust}.}
\begin{tabular}{lcc} \hline
\hline
& seed $M_{(i)}=10^{4}M_{\odot}$ & seed $M_{(ii)}=10^{6}M_{\odot}$  \\ 
\hline
$v^{r} = 0.0$ & $3.04 \times10^{7}M_{(i)}$   & $3.82\times10^{7}M_{(ii)}$\\
\hline
$v^{r} = -0.1$ & $2.49 \times10^{17}M_{(i)}$ & $2.07\times10^{16}M_{(ii)}$ \\
\hline
$v^{r} = -0.2$ &  $1.17\times10^{14}M_{(i)}$  & $3.28 \times10^{13}M_{(ii)}$  \\
\hline
$v^{r} = -0.3$ &  $1.96 \times10^{13}M_{(i)}$ & $7.82\times10^{12}M_{(ii)}$  \\
\hline
\hline
\end{tabular}
\end{table}


\subsection{Case $p \ne 0$}

In this case, the parameter space expands because now $\Gamma$ is a new parameter and so is the internal energy $\epsilon$ at initial time. In order to explore a wide parameter space we consider the following cases:

\begin{itemize}
\item[(a)] Again choose various values of the initial density $\rho=10^{-4},~10^{-6},~10^{-8},~10^{-10}$.
\item[(b)] We only study two values of the inward velocity, $v^r=-0.1,~-0.2$.
\item[(c)] We choose two values of the adiabatic constant $\Gamma = 1.1,~1.2$, so that we estimate self-interacting matter.
\item[(d)] And two values for the initial internal energy $\epsilon = 0.5,~1.0$, because unfortunately there is no a priori prescription for this parameter, unless for instance a polytropic equation of state is assumed initially, in which case the polytropic constant would be the free parameter.
\end{itemize}

\begin{figure*}
\includegraphics[width=7cm]{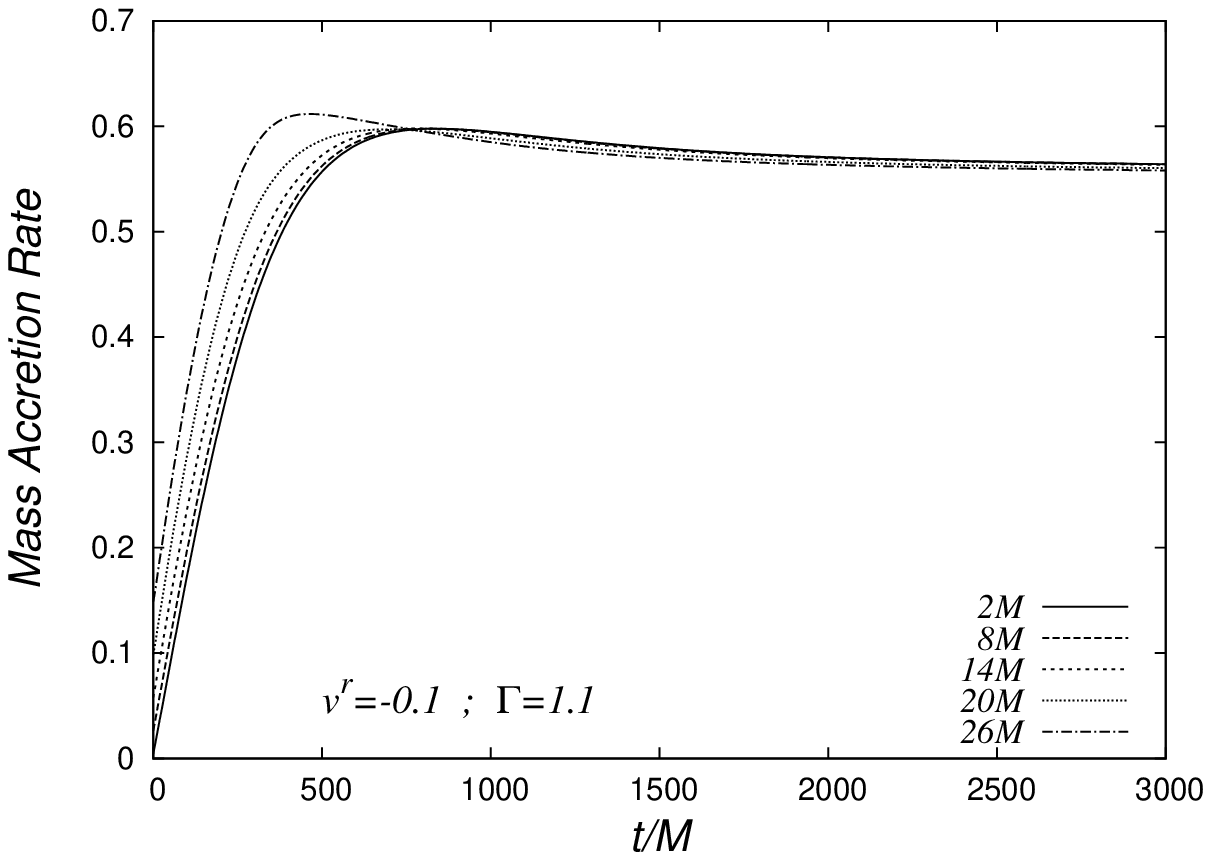}
\includegraphics[width=7cm]{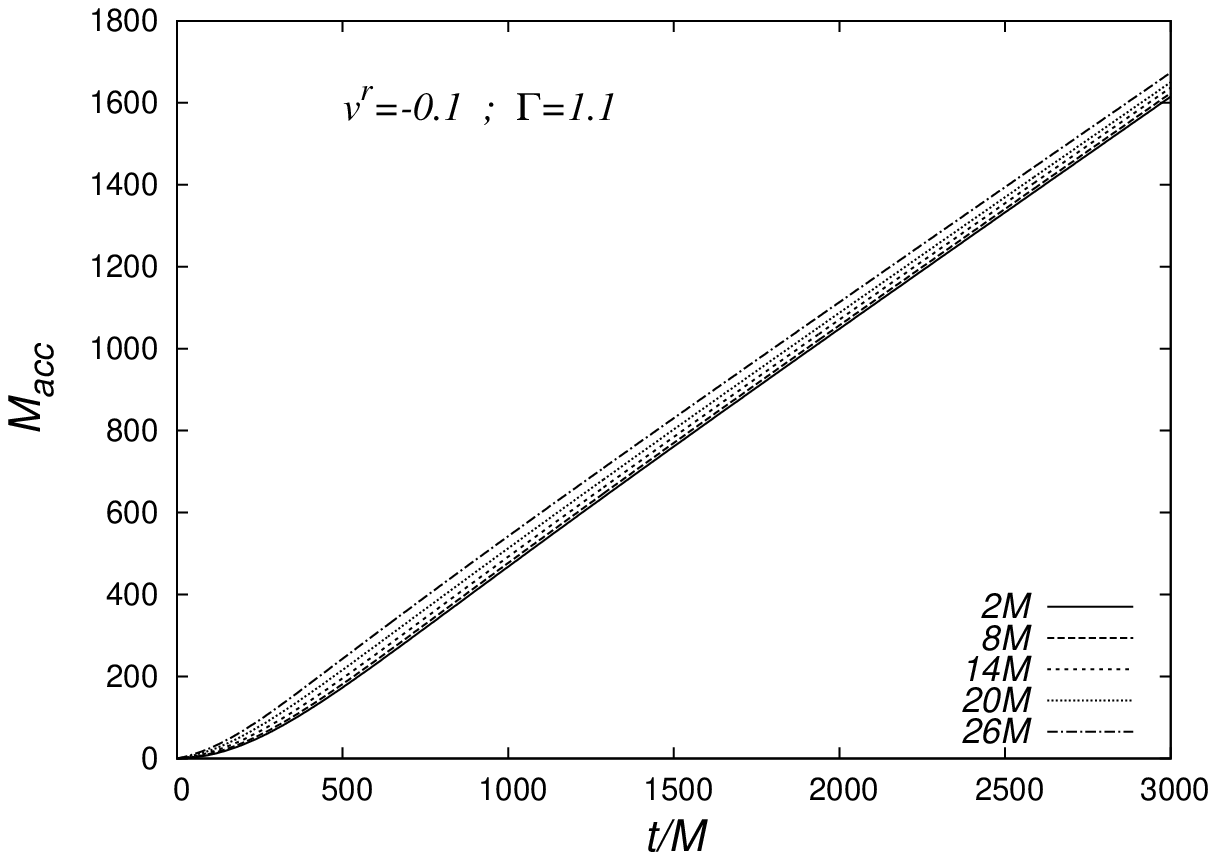}
\includegraphics[width=7cm]{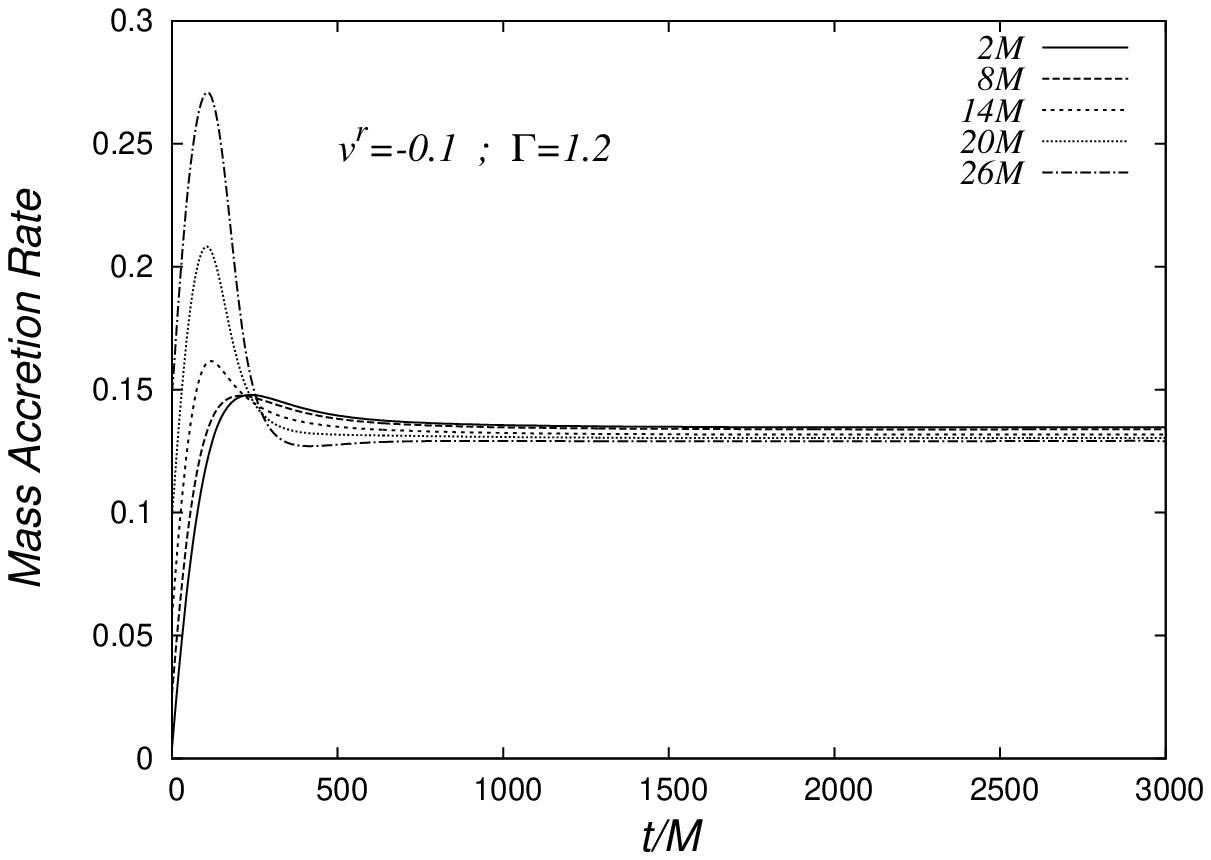}
\includegraphics[width=7cm]{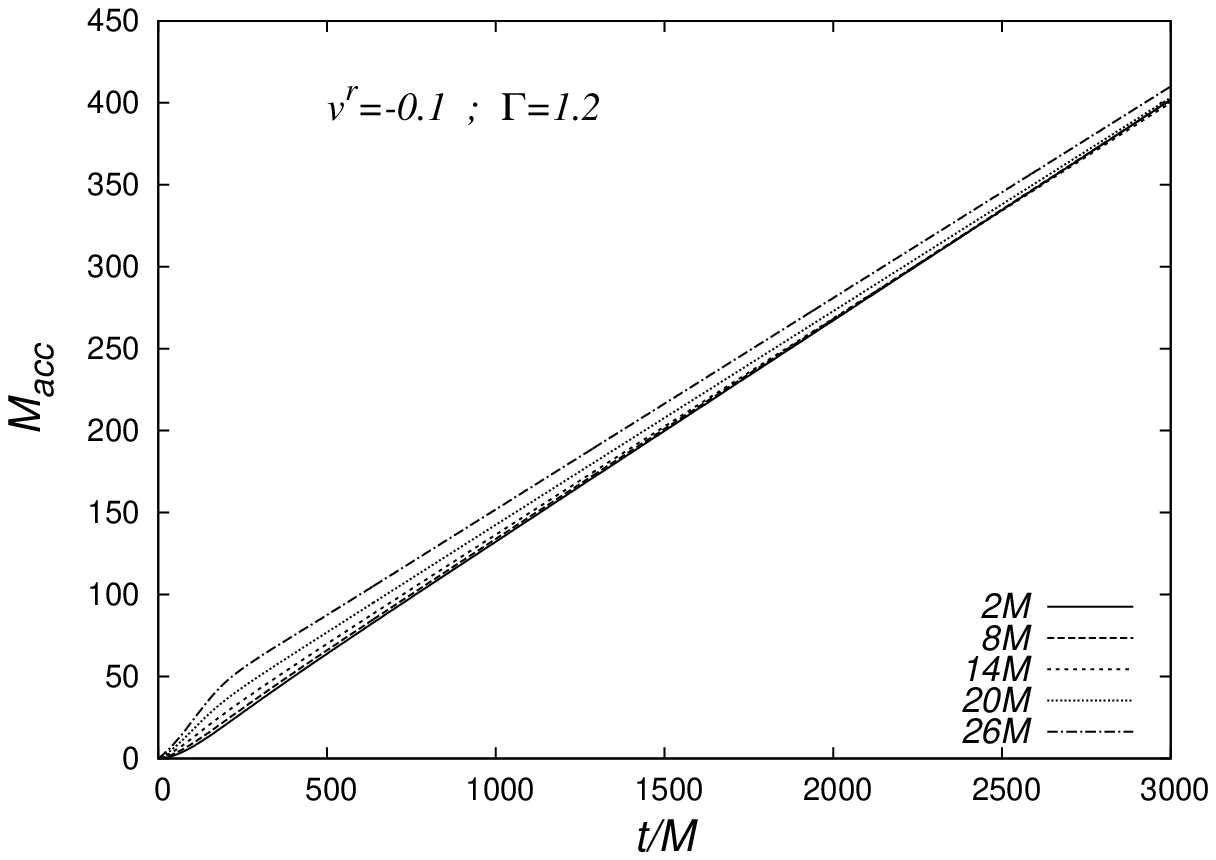}
\includegraphics[width=7cm]{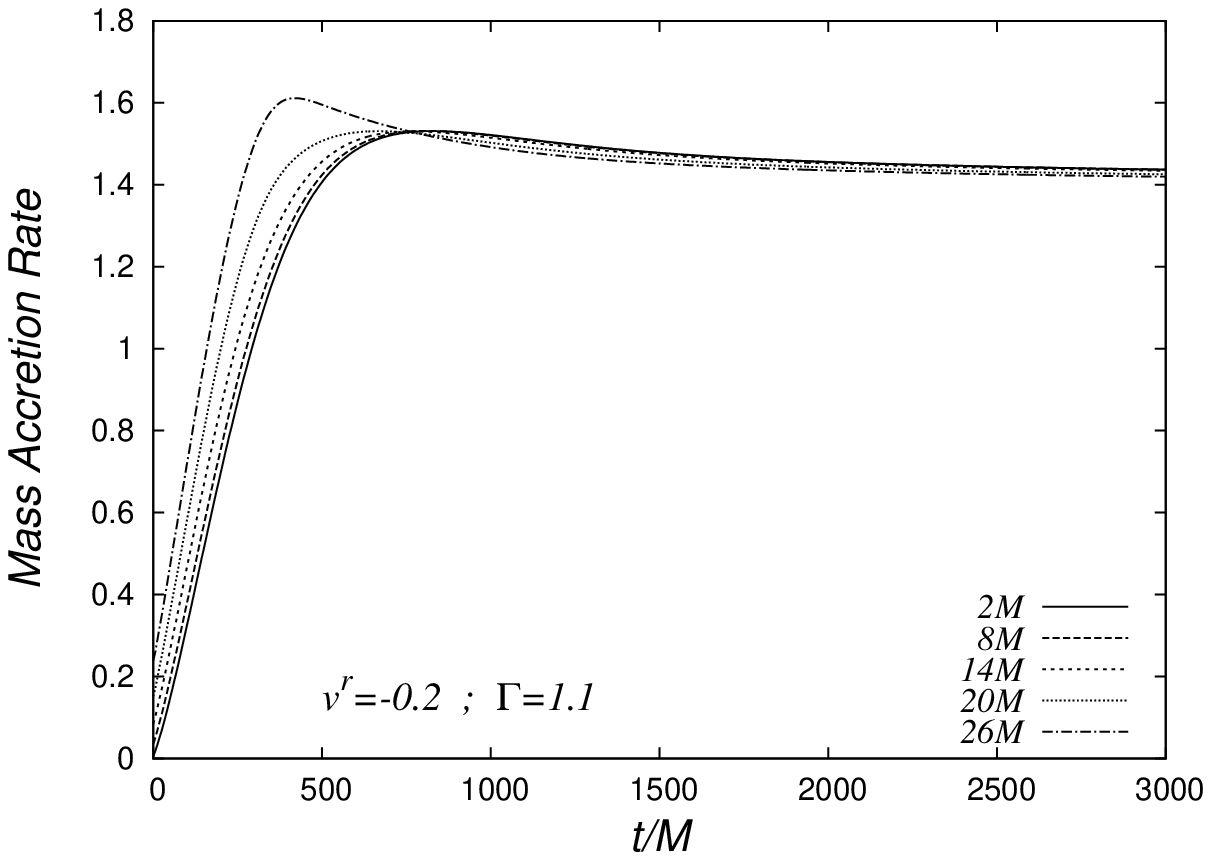}
\includegraphics[width=7cm]{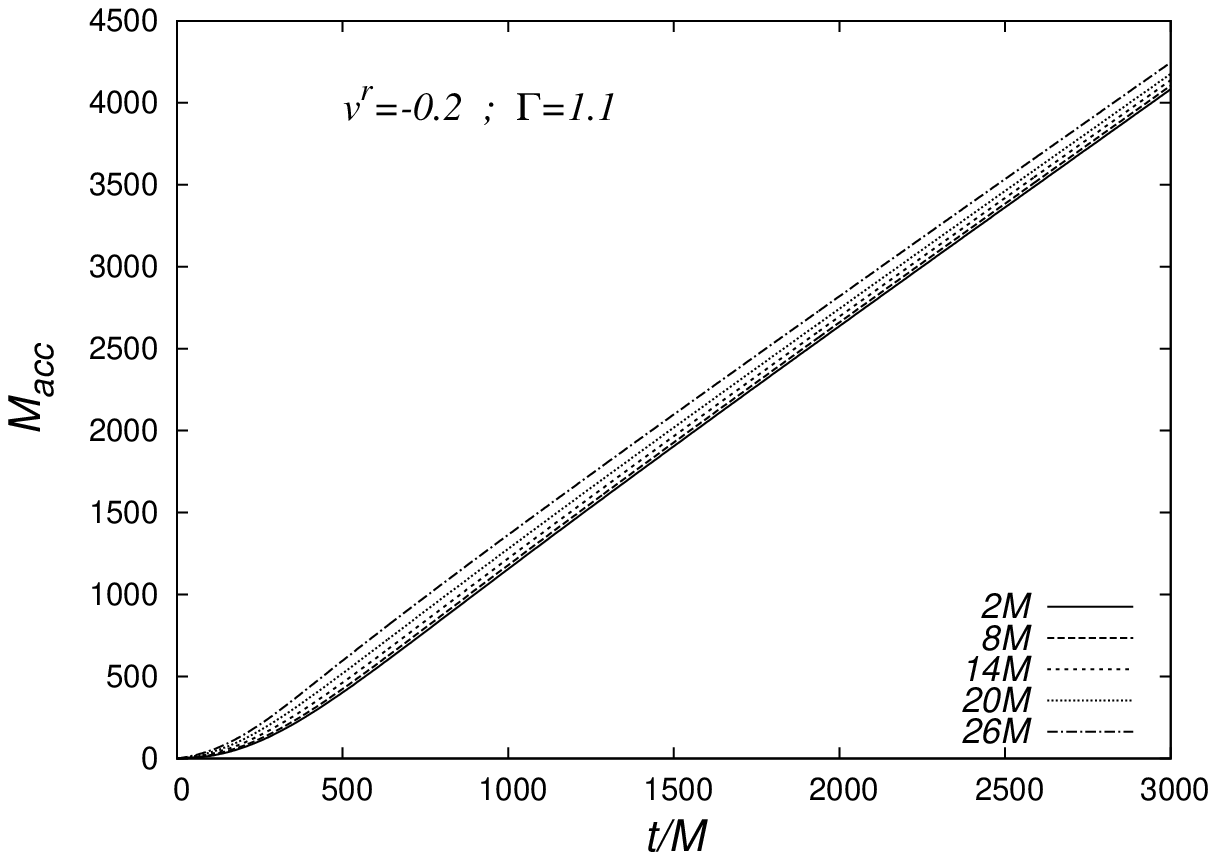}
\includegraphics[width=7cm]{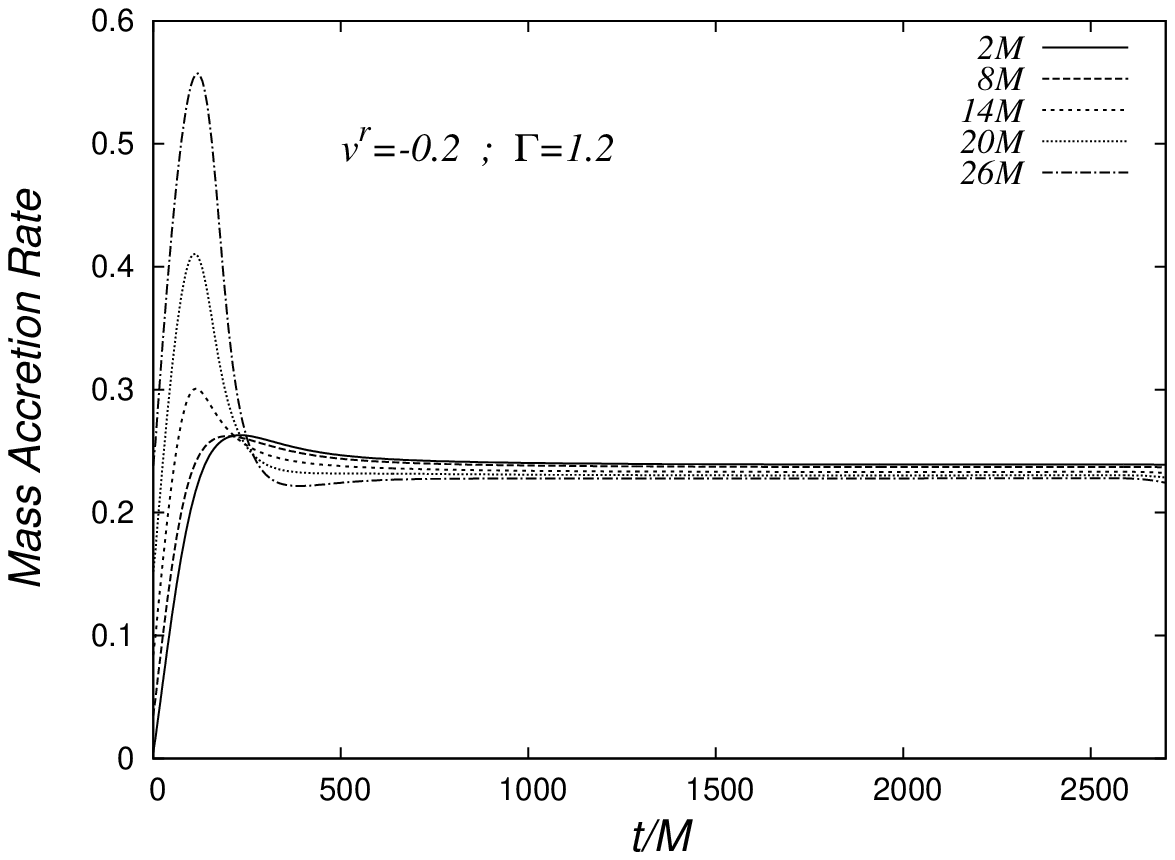}
\includegraphics[width=7cm]{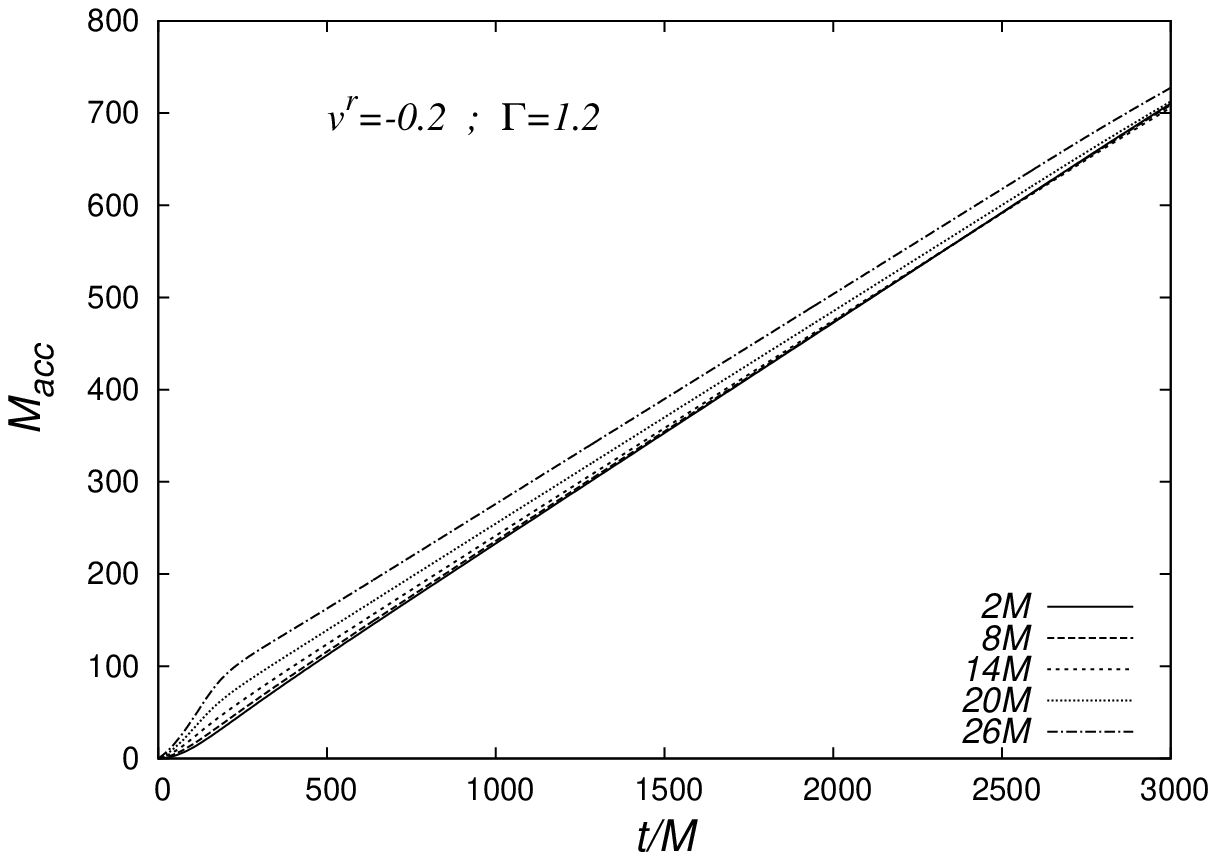}
\caption{\label{fig:MassAccretRateIdeal} In this figure we present the accretion rate of an ideal gas, for two values of the initial coordinate velocity $v^r$ and two values of the adiabatic constant $\Gamma$. In all the cases, the initial density and the specific internal energy used are $\rho=10^{-4}$ and $\epsilon=0.5$ respectively. The different type of line indicates that the accretion is being measured at various spherical surfaces, and it is remarkable that the accretion mass rate approaches the same value,  as happens in the $p=0$ case, however, in the present case, due to the non-zero pressure the result is non-trivial because outward and inward flow are expected to happen.}
\end{figure*}

\begin{figure}
\includegraphics[width=8cm]{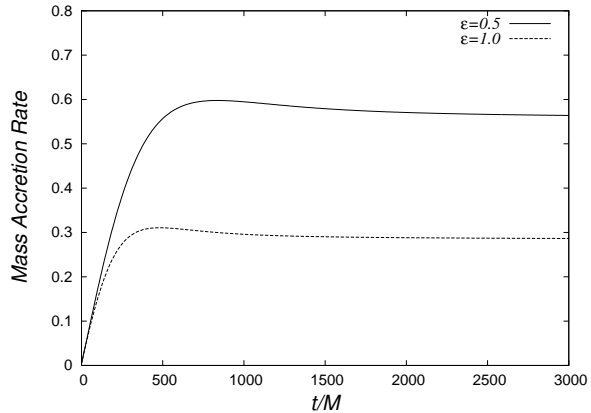}
\caption{\label{fig:energies} In this figure we present the accretion rate of an ideal gas, for two values of the initial  specific internal energy $\epsilon=0.5,~1.0$ and parameters: initial density $10^{-4}$, initial velocity $v^{r}=-0.1$ and adiabatic constant $\Gamma=1.1$. The role played by the internal energy at initial time consists in providing a value to the initial pressure, that is, when $\epsilon$ tends to zero the value of the pressure tends to zero too.}
\end{figure}

In all the combinations of these parameters we found that the accretion process reaches a stationary accretion mass rate. We show in Fig. \ref{fig:MassAccretRateIdeal} a representative sample of our results indicating a linear accreted mass in time after a transient lapse, where the initial relax toward a sort of stationary late-time attractor solution. For completeness in Fig. \ref{fig:energies} we show as an example the mass accretion rate with two different initial specific internal energies. This shows the role played by the initial internal energy, that is, in the limit of $\epsilon$ approaching zero as can be seen in (\ref{eq:eos}), the pressure tends to zero too and one recovers the unstable case.

Our results on the accretion mass rate and accreted mass are summarized in Table \ref{tab:MARIdeal1}, where, as in the pressure-less case, we also extrapolate our fitting results to the ultra low density corresponding to dark matter. The domain we use in these numerical experiments is $M<r<1001M$ and made sure the running time $t=3000M$ is adequate for this grid size and our code converges. We then use these extrapolated parameters to estimate the accreted mass during 10Gyr, which can be found in Table \ref{tab:MassTodayIdeal}.


\begin{table*}
\begin{minipage}{180mm}
\caption{\label{tab:MARIdeal1} In this table we show the parameters fitting the accreted mass for the initial internal energy  $\epsilon=0.5,~1.0$ and the adiabatic constant $\Gamma=1.1,~1.2$. We also show the extrapolation value corresponding to the dark matter density $\rho = 10^{-26}$ and $\rho=10^{-30}$ in geometric units, that correspond to two particular cases we analyze below. The anzat we use to fit the mass accreted is $M_{acc}=At^B + C$.}
\resizebox{7.0in}{!}{
\begin{tabular}{|l|l|l|llllllllllll} \hline
\multirow{1}{*}{} 
& $\rho=10^{-4}$ & $\rho=10^{-6}$ & $\rho=10^{-8}$ & $\rho=10^{-10}$ (Floor) & . . .& $\rho=10^{-26}$ & . . .& $\rho=10^{-30}$\\ \hline
\multicolumn{9}{|c|}{Extrapolating the accreted mass (Ideal Equation of state with $\Gamma=1.1$ and initial $\epsilon=0.5$)} \\ \hline
\multirow{3}{*}{$v^r = -0.1$} 
& $ A=0.808384 \pm 0.001781    $ & $ A=(0.808384 \pm 0.001781)*10^{-2} $ & $ A=(0.808384 \pm 0.001781)*10^{-4}   $ & $ A=(0.808384 \pm 0.001781)*10^{-6} $ & . . .& $ A=(0.808384 \pm 0.001781)*10^{-22}  $ & . . .& $ A=(0.808384 \pm 0.001781)*10^{-26} $ \\
& $ B=0.96003   \pm 0.0002526  $ & $ B=0.96003   \pm 0.0002526            $ & $ B=0.96003   \pm 0.0002526             $ & $ B=0.96003   \pm 0.0002526            $ & . . .& $ B=0.96003   \pm 0.0002526               $ & . . .& $ B=0.96003   \pm 0.0002526 $\\
& $ C=-145.048  \pm 0.3033       $ & $ C=(-145.048  \pm 0.3033)*10^{-2}    $ & $ C=(-145.048  \pm 0.3033)*10^{-4}      $ & $ C=(-145.048  \pm 0.3033)*10^{-6}    $ & . . .& $ C=(-145.048  \pm 0.3033)*10^{-22}     $ & . . .& $ C=(-145.048  \pm 0.3033)*10^{-26} $ \\
\hline
\multirow{3}{*}{$v^r = -0.2$}
& $ A=2.22404   \pm 0.006492    $ & $ A=(2.22404  \pm 0.006492)*10^{-2} $ & $ A=(2.22404   \pm 0.006492)*10^{-4}   $ & $ A=(2.22404   \pm 0.006492)*10^{-6} $ & . . .& $ A=(2.22404   \pm 0.006492)*10^{-22}  $ & . . .& $ A=(2.22404   \pm 0.006492)*10^{-26} $ \\
& $ B=0.951153  \pm 0.0003345  $ & $ B=0.951153  \pm 0.0003345           $ & $ B=0.951153  \pm 0.0003345            $ & $ B=0.951153  \pm 0.0003345            $ & . . .& $ B=0.951153  \pm 0.0003345               $ & . . .& $ B=0.951153  \pm 0.0003345 $\\
& $ C=-430.184   \pm 1.041        $ & $ C=(-430.184   \pm 1.041)*10^{-2}     $ & $ C=(-430.184   \pm 1.041)*10^{-4}      $ & $ C=(-430.184   \pm 1.041)*10^{-6}     $ & . . .& $ C=(-430.184   \pm 1.041)*10^{-22}      $ & . . .& $ C=(-430.184   \pm 1.041)*10^{-26} $ \\
\hline
\multicolumn{9}{|c|}{Extrapolating the accreted mass (Ideal Equation of state with $\Gamma=1.2$ and initial $\epsilon=0.5$)} \\ \hline
\multirow{3}{*}{$v^r = -0.1$} 
& $ A=0.145753 \pm 0.0002368   $ & $ A=(0.145753 \pm 0.0002368)*10^{-2} $ & $ A=(0.145753 \pm 0.0002368)*10^{-4}   $ & $ A=(0.145753 \pm 0.0002368)*10^{-6} $ & . . .& $ A=(0.145753 \pm 0.0002368)*10^{-22}  $ & . . .& $ A=(0.145753 \pm 0.0002368)*10^{-26} $ \\
& $ B=0.990879 \pm 0.0001922  $ & $ B=0.990879 \pm 0.0001922              $ & $ B=0.990879 \pm 0.0001922            $ & $ B=0.990879 \pm 0.0001922           $ & . . .& $ B=0.990879 \pm 0.0001922             $ & . . .& $ B=0.990879 \pm 0.0001922 $\\
& $ C=-4.70678  \pm 0.04092      $ & $ C=(-4.70678  \pm 0.04092)*10^{-2}    $ & $ C=(-4.70678  \pm 0.04092)*10^{-4}      $ & $ C=(-4.70678  \pm 0.04092)*10^{-6}    $ & . . .& $ C=(-4.70678  \pm 0.04092)*10^{-22}     $ & . . .& $ C=(-4.70678  \pm 0.04092)*10^{-26} $ \\
\hline
\multirow{3}{*}{$v^r = -0.2$}
& $ A=0.260364  \pm 0.0004783    $ & $ A=(0.260364  \pm 0.0004783)*10^{-2} $ & $ A=(0.260364  \pm 0.0004783)*10^{-4}   $ & $ A=(0.260364  \pm 0.0004783)*10^{-6} $ & . . .& $ A=(0.260364  \pm 0.0004783)*10^{-22}  $ & . . .& $ A=(0.260364  \pm 0.0004783)*10^{-26} $ \\
& $ B=0.990018  \pm 0.0002173  $ & $ B=0.990018  \pm 0.0002173          $ & $ B=0.990018  \pm 0.0002173          $ & $ B=0.990018  \pm 0.0002173            $ & . . .& $ B=0.990018  \pm 0.0002173               $ & . . .& $ B=0.990018  \pm 0.0002173 $ \\
& $ C=-10.0288   \pm 0.08219        $ & $ C=(-10.0288   \pm 0.08219)*10^{-2}     $ & $ C=(-10.0288   \pm 0.08219)*10^{-4}      $ & $ C=(-10.0288   \pm 0.08219)*10^{-6}     $ & . . .& $ C=(-10.0288   \pm 0.08219)*10^{-22}      $ & . . .& $ C=(-10.0288   \pm 0.08219)*10^{-26} $ \\
\hline
\multicolumn{9}{|c|}{Extrapolating the accreted mass (Ideal Equation of state with $\Gamma=1.1$ and initial $\epsilon=1.0$)} \\ \hline
\multirow{3}{*}{$v^r = -0.1$} 
& $ A=0.396444 \pm 0.001466     $ & $ A=(0.396444 \pm 0.001466)*10^{-2} $ & $ A=(0.396444 \pm 0.001466)*10^{-4}   $ & $ A=(0.396444 \pm 0.001466)*10^{-6} $ & . . .& $ A=(0.396444 \pm 0.001466)*10^{-22}  $ & . . .& $ A=(0.396444 \pm 0.001466)*10^{-26} $ \\
& $ B=0.963098   \pm 0.0004361  $ & $ B=0.963098   \pm 0.0004361            $ & $ B=0.963098   \pm 0.0004361             $ & $ B=0.963098   \pm 0.0004361            $ & . . .& $ B=0.963098   \pm 0.0004361              $ & . . .& $ B=0.963098   \pm 0.0004361 $\\
& $ C=-39.3563  \pm 0.2112        $ & $ C=(-39.3563  \pm 0.2112)*10^{-2}    $ & $ C=(-39.3563  \pm 0.2112)*10^{-4}      $ & $ C=(-39.3563  \pm 0.2112)*10^{-6}    $ & . . .& $ C=(-39.3563  \pm 0.2112)*10^{-22}     $ & . . .& $ C=(-39.3563  \pm 0.2112)*10^{-26} $ \\
\hline
\multirow{3}{*}{$v^r = -0.2$}
& $ A=0.891524   \pm 0.003389    $ & $ A=(0.891524   \pm 0.003389)*10^{-2} $ & $ A=(0.891524   \pm 0.003389)*10^{-4}   $ & $ A=(0.891524   \pm 0.003389)*10^{-6} $ & . . .& $ A=(0.891524   \pm 0.003389)*10^{-22}  $ & . . .& $ A=(0.891524   \pm 0.003389)*10^{-26} $ \\
& $ B=0.961579  \pm 0.0004483  $ & $ B=0.961579  \pm 0.0004483           $ & $ B=0.961579  \pm 0.0004483           $ & $ B=0.961579  \pm 0.0004483            $ & . . .& $ B=0.961579  \pm 0.0004483               $ & . . .& $ B=0.961579  \pm 0.0004483 $\\
& $ C=-100.255   \pm 0.4835        $ & $ C=(-100.255   \pm 0.4835)*10^{-2}     $ & $ C=(-100.255   \pm 0.4835)*10^{-4}      $ & $ C=(-100.255   \pm 0.4835)*10^{-6}     $ & . . .& $ C=(-100.255   \pm 0.4835)*10^{-22}      $ & . . .& $ C=(-100.255   \pm 0.4835)*10^{-26} $ \\
\hline
\multicolumn{9}{|c|}{Extrapolating the accreted mass (Ideal Equation of state with $\Gamma=1.2$ and initial $\epsilon=1.0$)} \\ \hline
\multirow{3}{*}{$v^r = -0.1$} 
& $ A=0.0891933 \pm 0.0003314   $ & $ A=(0.0891933 \pm 0.0003314)*10^{-2} $ & $ A=(0.0891933 \pm 0.0003314)*10^{-4}   $ & $ A=(0.0891933 \pm 0.0003314)*10^{-6} $ & . . .& $ A=(0.0891933 \pm 0.0003314)*10^{-22}  $ & . . .& $ A=(0.0891933 \pm 0.0003314)*10^{-26} $ \\
& $ B=0.987067 \pm 0.0004521  $ & $ B=0.987067 \pm 0.0004521              $ & $ B=0.987067 \pm 0.0004521          $ & $ B=0.987067 \pm 0.0004521           $ & . . .& $ B=0.987067 \pm 0.0004521             $ & . . .& $ B=0.987067 \pm 0.0004521$\\
& $ C=-1.86548  \pm 0.04307      $ & $ C=(-1.86548  \pm 0.04307)*10^{-2}    $ & $ C=(-1.86548  \pm 0.04307)*10^{-4}      $ & $ C=(-1.86548  \pm 0.04307)*10^{-6}    $ & . . .& $ C=(-1.86548  \pm 0.04307)*10^{-22}     $ & . . .& $ C=(-1.86548  \pm 0.04307)*10^{-26} $ \\
\hline
\multirow{3}{*}{$v^r = -0.2$}
& $ A=0.135829  \pm 0.0001597    $ & $ A=(0.135829  \pm 0.0001597)*10^{-2} $ & $ A=(0.135829  \pm 0.0001597)*10^{-4}   $ & $ A=(0.135829  \pm 0.0001597)*10^{-6} $ & . . .& $ A=(0.135829  \pm 0.0001597)*10^{-22}  $ & . . .& $ A=(0.135829  \pm 0.0001597)*10^{-26} $ \\
& $ B=0.997385  \pm 0.0001416  $ & $ B=0.997385  \pm 0.0001416          $ & $ B=0.997385  \pm 0.0001416          $ & $ B=0.997385  \pm 0.0001416            $ & . . .& $ B=0.997385  \pm 0.0001416               $ & . . .& $ B=0.997385  \pm 0.0001416 $ \\
& $ C=-1.54871   \pm 0.02557        $ & $ C=(-1.54871   \pm 0.02557)*10^{-2}     $ & $ C=(-1.54871   \pm 0.02557)*10^{-4}      $ & $ C=(-1.54871   \pm 0.02557)*10^{-6}     $ & . . .& $ C=(-1.54871   \pm 0.02557)*10^{-22}      $ & . . .& $ C=(-1.54871   \pm 0.02557)*10^{-26} $ \\
\hline
\end{tabular} }
\end{minipage}
\end{table*}

\begin{figure}
\includegraphics[width=8cm]{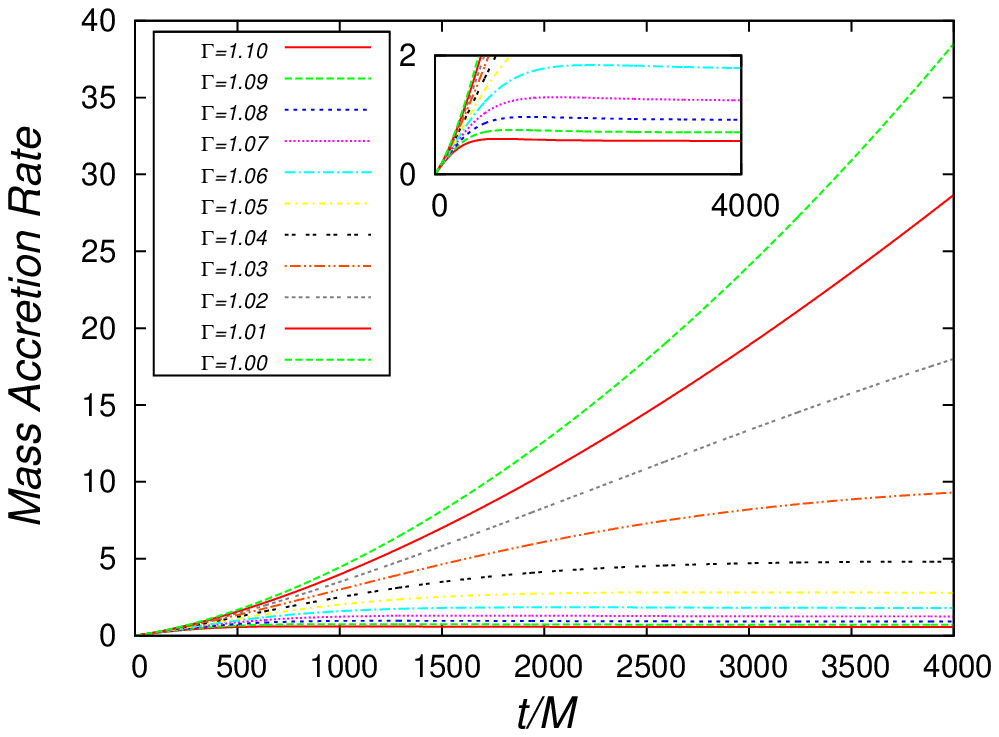}
\includegraphics[width=8cm]{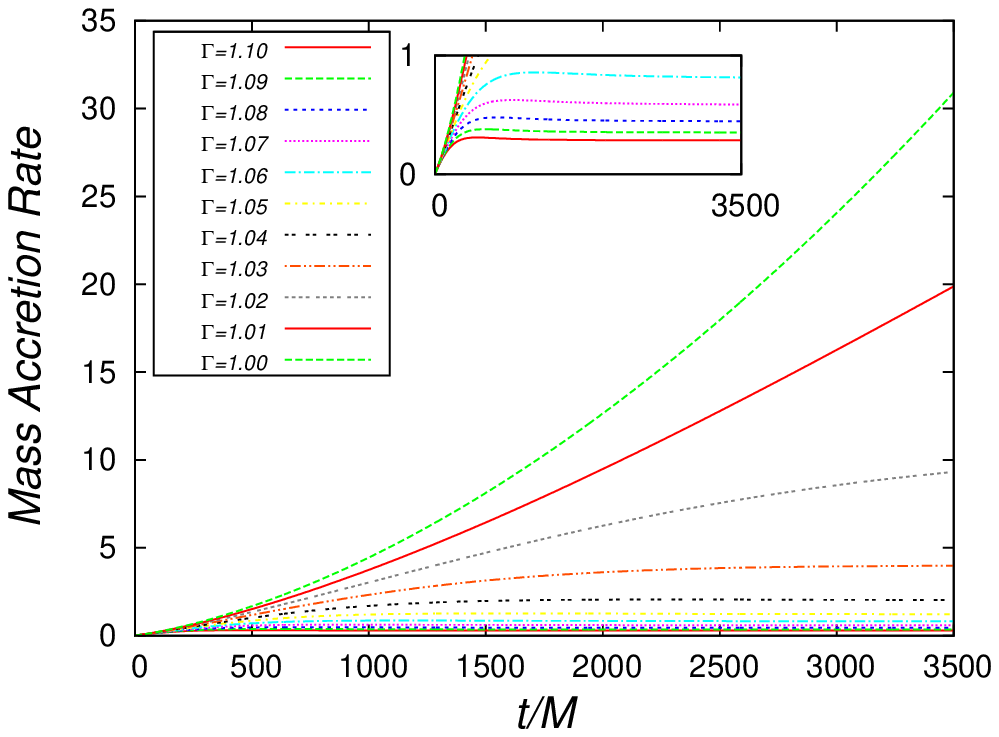}
\caption{\label{fig:gammas} We show the accretion mass rate for various values of $\Gamma$ for the configurations corresponding to the initial parameters $\rho=10^{-4}$, $v^r = -0.1$, $\epsilon = 0.5, 1$. In the insets we show a zoom in of the accretion mass rate that illustrates the various behaviors at the initial transient stage that depends on $\Gamma$.}
\end{figure}

From our results it call the attention that approaching a stationary accretion process is very sensitive to the value of $\Gamma$, as can be seen in the Tables \ref{tab:MassToday} and  \ref{tab:MassTodayIdeal}, where the resulting estimates in the accreted mass can change by about fifteen orders of magnitude when changing $\Gamma$ from 1 to 1.1, which corresponds to the system of dust and our ideal gas dark matter model with pressure respectively. In order to show that this is not an artifact of our numerical approach, we show in Fig. \ref{fig:gammas} that the accretion mass rate changes monotonically with $\Gamma$ for two configurations chosen from the collection of cases we experienced with.

\begin{table}
\caption{\label{tab:MassTodayIdeal} In this table, we show the mass accreted by the a black hole since 10Gys ago to the present for two black hole seeds. The initial density profile of the dark matter is $\rho=100 M_{\odot} pc^{-3}$, which in geometric units, for black hole seed masses  $M_{(i)}=10^{4}M_{\odot}$ and $M_{(ii)}=10^{6}M_{\odot}$,  correspond to $\rho = 1.17\times10^{-30}$ and $\rho = 1.17\times10^{-26}$ in geometric units respectively, which for the density corresponds to $1/lenght^2$, which in our case the unit length is $M$ . The values of the mass accreted for these densities are taken from the extrapolation in Table \ref{tab:MARIdeal1}.}
\begin{tabular}{lcc} \hline
\hline
& seed $M_{(i)}=10^{4}M_{\odot}$ & seed $M_{(ii)}=10^{6}M_{\odot}$  \\ 
\hline
\hline
& $\Gamma = 1.1$ , $\epsilon=0.5$ & \\
\hline
\hline
$v^{r} = -0.1$ & $9.21\times10^{-9}M_{(i)}$ & $1.11\times10^{-6}M_{(ii)}$\\
\hline
$v^{r} = -0.2$ & $1.72\times10^{-8}M_{(i)}$ & $2.16\times10^{-6}M_{(ii)}$ \\
\hline
\hline
& $\Gamma = 1.2$ , $\epsilon=0.5$ & \\
\hline
\hline
$v^{r} = -0.1$ & $ 6.31\times10^{-9}M_{(i)}$   &  $6.59\times10^{-7}M_{(ii)}$ \\
\hline
$v^{r} = -0.2$ & $1.1\times10^{-8}M_{(i)}$ & $1.14\times10^{-6}M_{(ii)}$   \\
\hline
\hline
& $\Gamma = 1.1$ , $\epsilon=1.0$ & \\
\hline
\hline
$v^{r} = -0.1$ & $5.2\times10^{-9}M_{(i)}$ & $6.11\times10^{-7}M_{(ii)}$  \\
\hline
$v^{r} = -0.2$ & $1.1\times10^{-8}M_{(i)}$ & $1.29\times10^{-6}M_{(ii)}$ \\
\hline
\hline
& $\Gamma = 1.2$ , $\epsilon=1.0$ &   \\
\hline
\hline
$v^{r} = -0.1$ & $3.28\times10^{-9}M_{(i)}$  & $3.48\times10^{-7}M_{(ii)}$ \\
\hline
$v^{r} = -0.2$ &   $7.80\times10^{-9}M_{(i)}$ &  $7.90\times10^{-7}M_{(ii)}$ \\
\hline
\hline
\end{tabular}
\end{table}

\begin{figure}
\includegraphics[width=8cm]{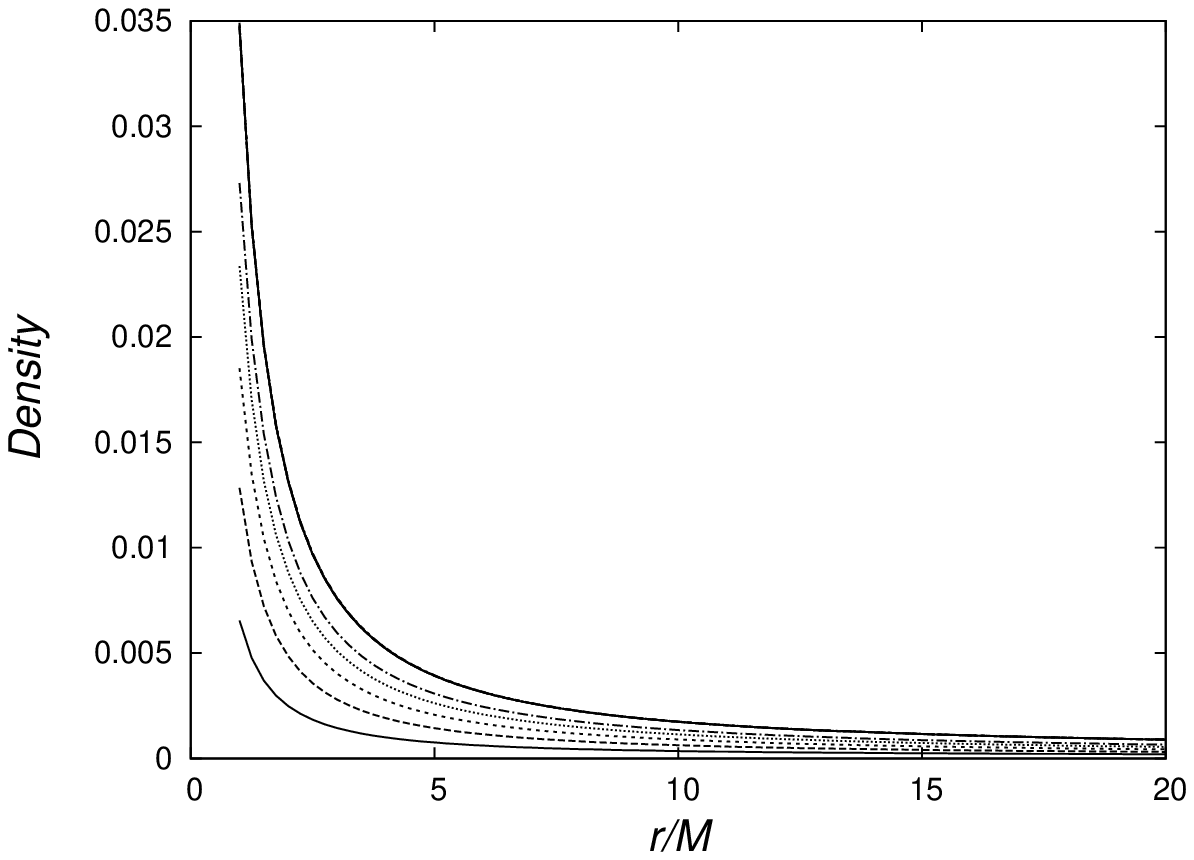}
\includegraphics[width=8cm]{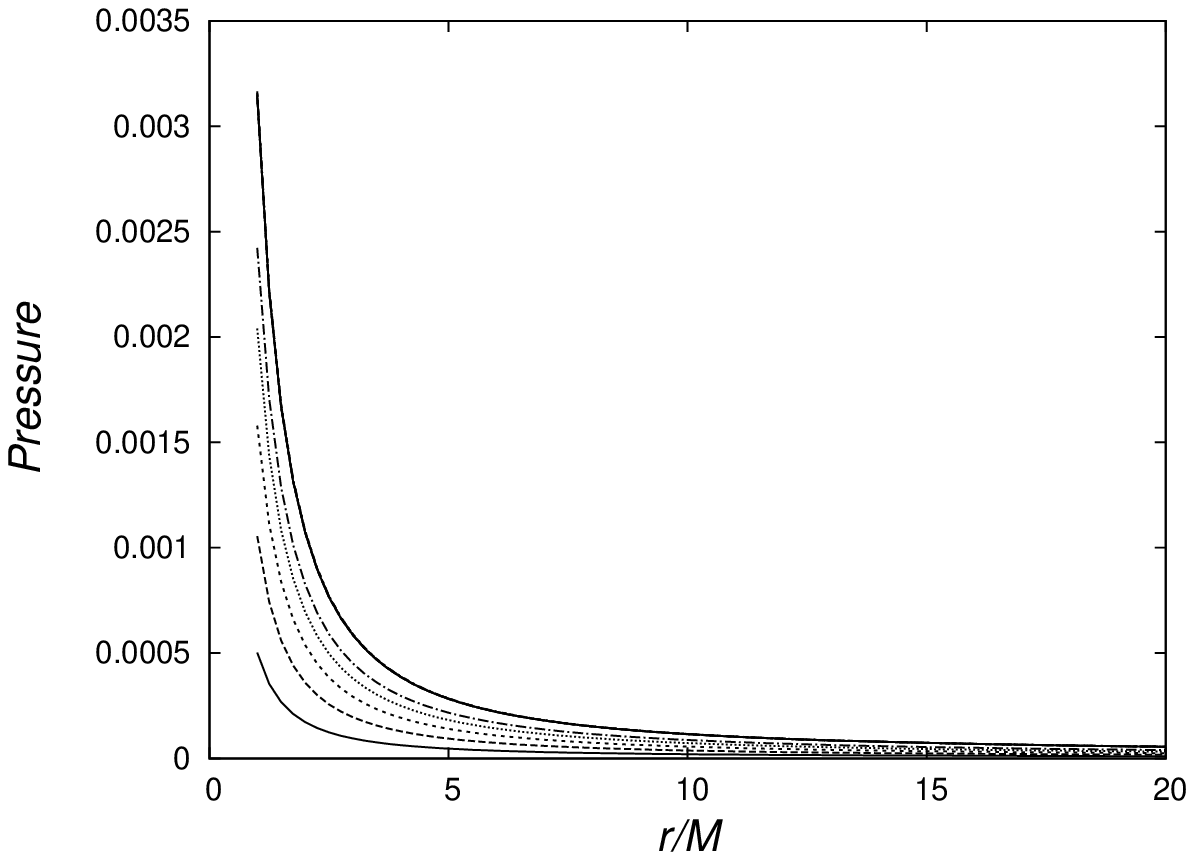}
\includegraphics[width=8cm]{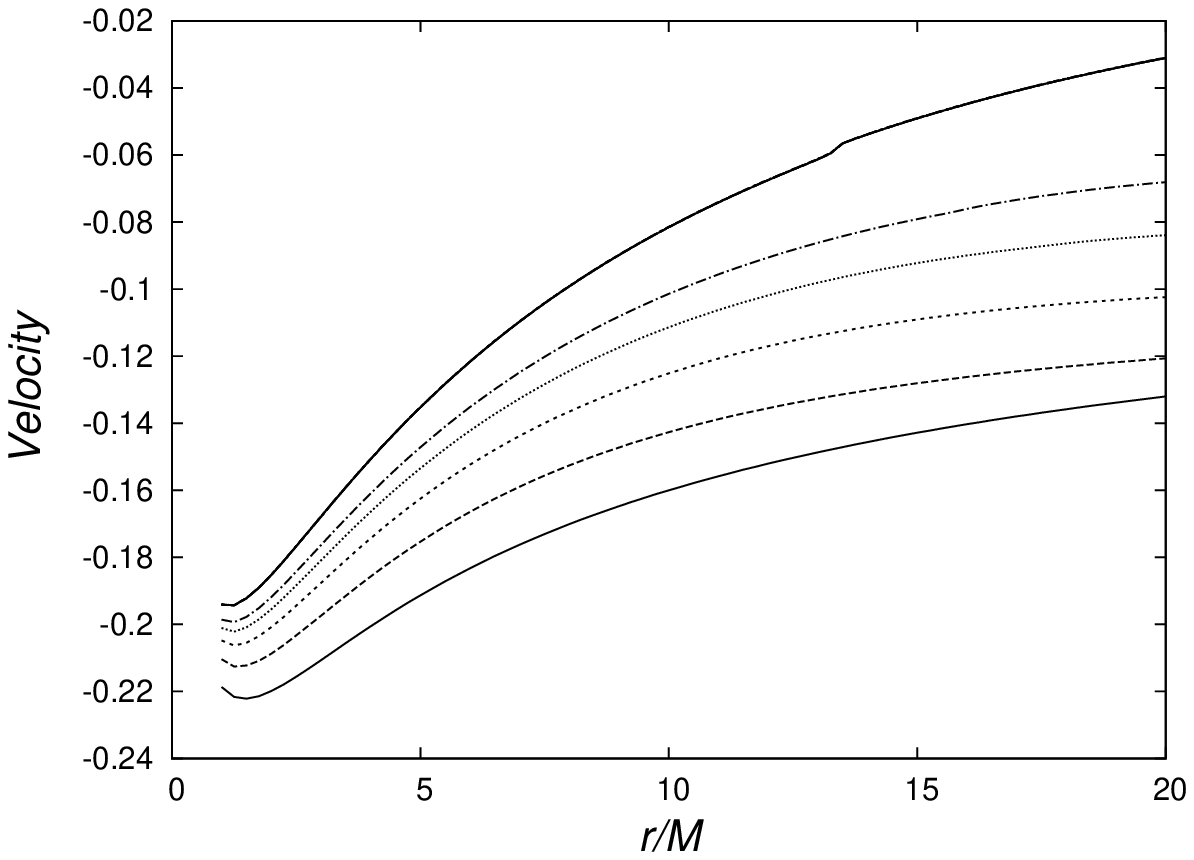}
\caption{\label{fig:steady} In this figure we show how after an initial transient the state variables approach a steady state, which provokes the accretion mass rate to be stationary. This case corresponds to the initial parameters $\rho=10^{-4}$, $v^r=-0.1$, $\Gamma=1.1$ and $\epsilon=0.5$. The set of  gray, dashed and dotted lines correspond to snapshots at early times during the simulation, whereas the black bold line is actually a set of various snapshots superposed after the system enters a steady stage. The curves start from the excision boundary at $r=M$ and we only show the region near the horizon where these quantities have structure.}
\end{figure}

{\it Long time attractor behavior of the solutions.} 
As an example of how the state variables approach a stationary regime in the case of non zero pressure, we show in Fig. \ref{fig:steady} snapshots of $\rho$, $p$ and $v^r$. It can be noticed that the intial data corresponding to constant density, pressure and velocity, after a transient lapse approach a time-independent regime. It can also be observed that the pressure is bigger near the horizon -because it is linear with the rest mass density- which prevents the fluid from falling quickly into the hole, an effect that does not happen in the $p=0$ case and we consider to be the responsible of slowing down the accretion rates.

\subsection{Test of the code}

\begin{figure}
\includegraphics[width=8cm]{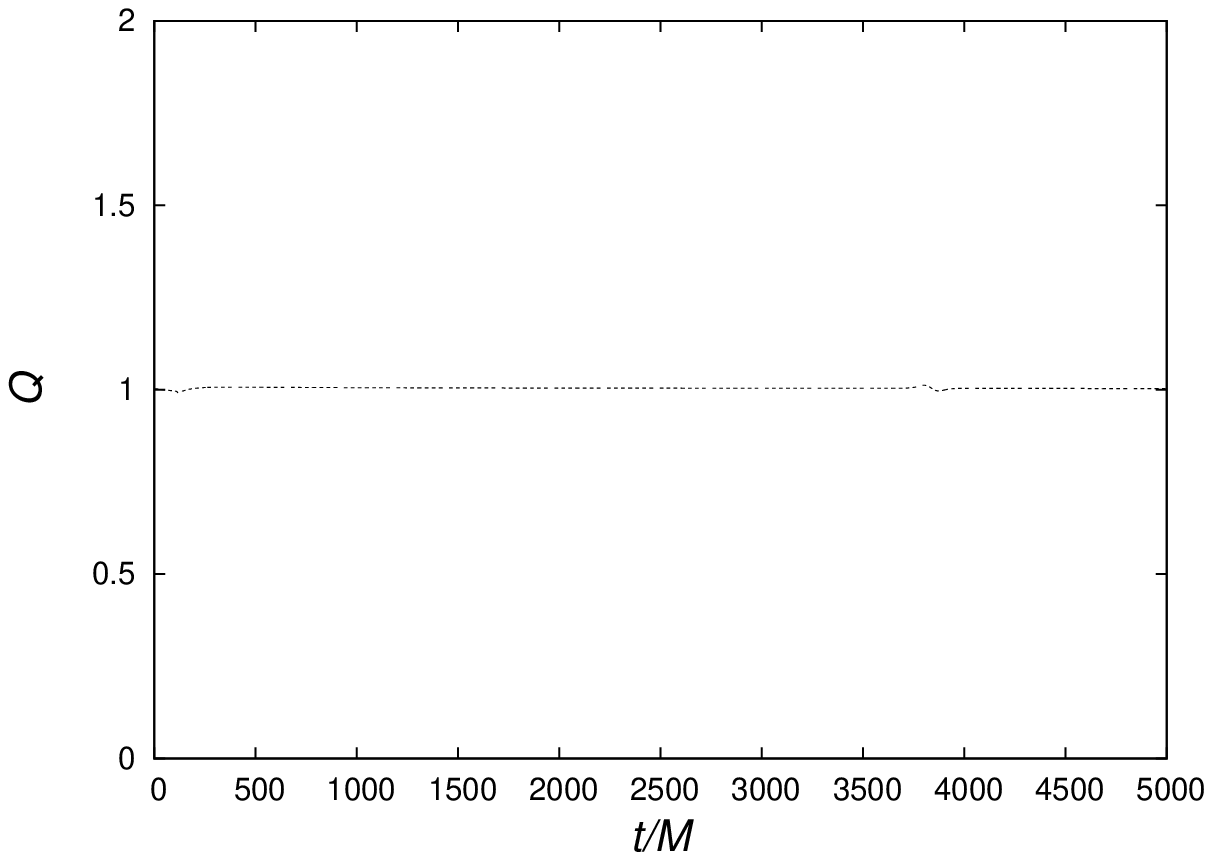} 
\includegraphics[width=8cm]{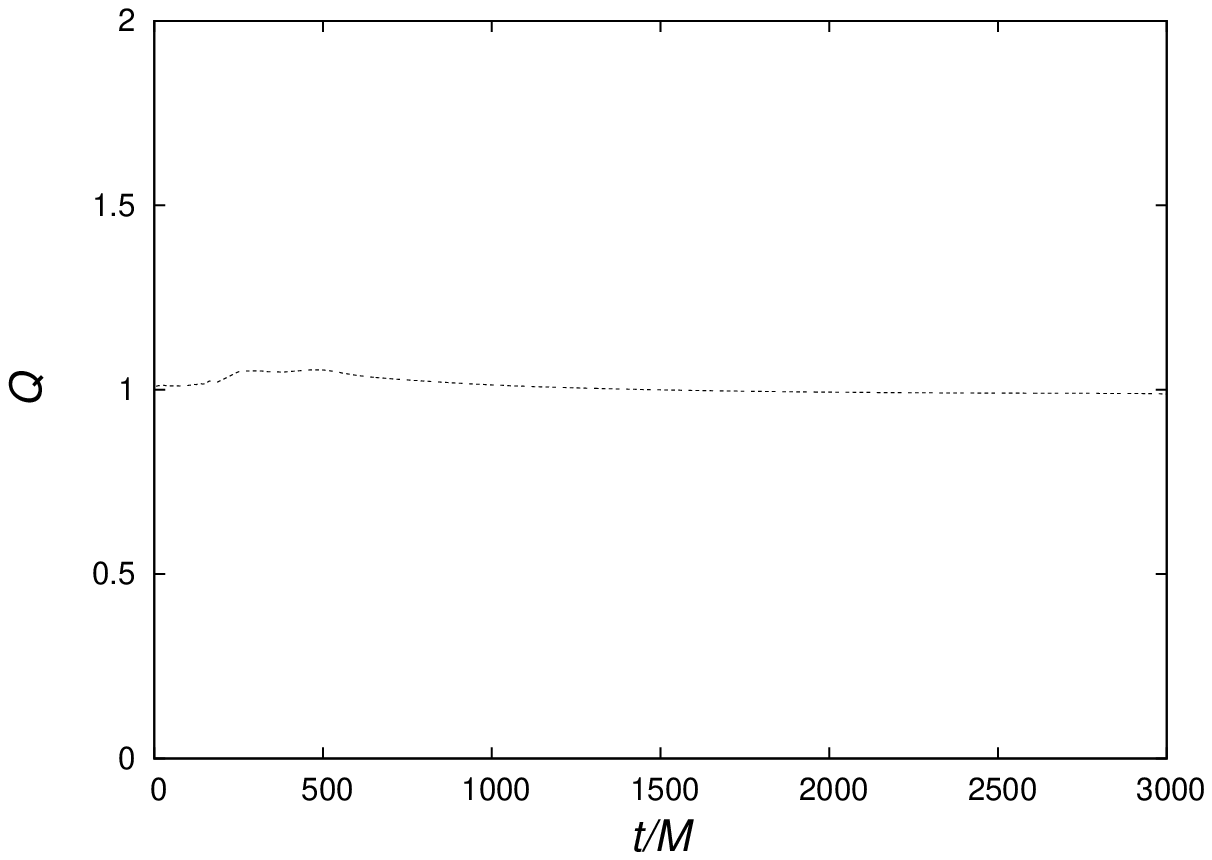} 
\caption{\label{fig:convergence} In this figure we show the order of self-convergence of $\rho$, for two different physical cases, 1) (left panel) $p=0$, $\rho=10^{-4}$ and $v^r=-0.1$ and, 2) (right panel) $\Gamma=1.2$, $\rho=10^{-4}$ $v^r=-0.1$, $\epsilon=0.5$. We calculate the convergence using the $L_1$ norm of the differences between the value of the density for three different resolutions $\Delta x_1=0.1$, $\Delta x_2=\Delta x_1/2$ and  $\Delta x_3=\Delta x_2/2$.  Then, the self-convergece factor, $Q$, for these three resolutions is calculated from the following expresion as  $2^{Q_{sc}}=\Delta E_1/\Delta E_2$ where $\Delta E_1 = L_1(\rho_1 - \rho_2)$ and $\Delta E_2 = L_1(\rho_2 - \rho_3)$. The dominant approximation is that of the reconstruction of variables at the interface cells, which is a piece wise constant and therefore first order accurate. This is consistent with the self-convergence of this plot.}
\end{figure}

In order to validate our numerical results we present a very stringent test which is related to the  convergence of our numerical solution. Since we do not have an exact solution of the equations, we perform a Cauchy type of self-convergence test on the state variables. As a sample of such test we show in Fig. \ref{fig:convergence} the self-convergence of the rest mass density for two of our runs. We find that the order of self-convergence is one, which is consistent with the order of approximation (first order) of our variable reconstruction at the cell interfaces of the Riemann solver.

\section{Conclusions}
\label{sec:conclusions}

We have numerically studied the spherical accretion of a perfect fluid with an ideal gas equation of state on a fixed Schwarzschild background space-time in penetrating coordinates and support our results with self-convergence tests. Our procedure consists in the numerical solution of the time-dependent relativistic Euler equations on the black hole space-time. As a first approximation, due to well known numerical difficulties, we have extrapolated our results to the physical values corresponding to the accretion of dark matter onto a supermassive black hole. We have explored a wide range of parameters related to density, pressure, internal energy, adiabatic constant and inward initial velocity of the fluid.

For the parameter space explored, we have found that when the pressure of the dark matter is zero, within the window of our simulations there is no steady state accretion process, and also that the current masses of supermassive black holes should be orders of magnitude bigger than those observed if they are assumed to be fed by spherical accretion of dark matter. 
However, for the case of non-zero pressure, we have found in all the experiments we have analyzed that a steady state accretion is reached and behaves like a late-time attractor solution which is reached sooner for bigger values of the adiabatic index.

We found that for initial black hole seed of already supermassive size ($10^6M_{\odot}$), the accreted mass during 10Gyr is of the order of solar masses. Either this implies that dark matter may obey an equation of state with non-zero pressure, or in a conservative case, the accretion of dark matter onto supermassive black holes contributes with a small fraction compared to possible accretion of baryonic matter as suggested by the bolometric luminosity of quasars \cite{Peirani2008}, where a bound of 10\% of the accretion is allowed to be dark matter.

This spherically symmetric accretion model is a first step toward a full study of the SMBH plus dark matter system, which in a complete version should involve the introduction of tidal forces due to angular components of the fluid that will break the spherical symmetry, back reaction effects or full non-linear evolution of the black hole,  transfer of angular momentum to the black hole, etc. 




\section*{Acknowledgments}

This research is partly supported by grants:
CIC-UMSNH-4.9 and
CONACyT 106466.                                               
The runs were carried out in the IFM Cluster.





\label{lastpage}

\end{document}